\documentclass{article}
\usepackage{epsfig}

\date{\today}

\begin{document}
\begin{center}
{\Large\bf Gravitating nonabelian solutions with NUT charge }
\vspace{0.6cm}
\\
Eugen Radu
\\
{\small
\it Albert-Ludwigs-Universit\"at Freiburg, 
\\Fakult\"at f\"ur Mathematik und Physik,
Physikalisches Institut,
\\Hermann-Herder-Stra\ss e 3, D-79104 Freiburg, Germany
}
\end{center}
\begin{abstract}
We argue that
the Einstein-Yang-Mills theory
presents nontrivial solutions with a NUT charge.
These solutions approach asymptotically the Taub-NUT spacetime. 
They are characterized by the NUT parameter, the mass and the 
node numbers of the magnetic potential and present 
both electric and magnetic potentials.
 
The existence of nontrivial Einstein-Yang-Mills solutions with NUT charge
in the presence of a negative cosmological constant is also discussed.
We use the counterterm subtraction method to calculate the boundary energy-momentum
tensor and the mass of these configurations. Also, dyon black hole solutions with
nonspherical event horizon topology  are shown to exist for a negative cosmological
constant.

\end{abstract}
\vspace{0.5cm}
\section{Introduction}
According to the so called "no-hair" conjecture, an asymptotically flat,
stationary black hole is uniquely
described in terms of a small set of asymptotically measurable quantities.
It was believed that this property of black holes, following from the uniqueness
and nonexistence results proven rigorously for a variety of field theories,
persists for general matter sources. The hypothesis was disproved more than 
ten years ago, 
when several authors presented a counterexample within the framework of $SU(2)$ 
Einstein-Yang-Mills (EYM) theory \cite{89}.
Although the new solution was static and had vanishing Yang-Mills (YM) 
charges, it was different from the Schwarzschild black hole and,
therefore, not characterized by its total mass. 
Despite its lack of stability, this nonabelian solution still
presents a challenge to the no hair conjecture.
Since then, a lot of black hole with different hairs have been found
(for a general review of such solutions see \cite{Volkov:1999cc,Gal'tsov:2001tx}).

It is however worth inquiring, what happens if we drop asymptotic flatness?
Will these "hairy" black holes survive?
As proven by various authors, this is the case in the presence of a 
cosmological constant $\Lambda$.
In the last years, nonabelian black hole solutions 
in a cosmological context were considered by various authors.
Remarkably enough, for an asymptotically anti-de Sitter (AdS) spacetime,
there are even stable solutions \cite{Winstanley:1999sn}.

As discussed in \cite{Winstanley:1999sn, Bjoraker:2000qd}, 
a variety of well known features of asymptotically self-gravitating nonabelian
solutions are not shared by their AdS counterparts. 
For example, there are finite energy solutions for 
several intervals of the shooting parameter 
(the value of gauge function at the event horizon), rather than discrete values. 
There are also stable monopole solutions in which the gauge field has no zeros, and the asymptotic
value of the gauge potential is not fixed.
This behavior is drastically different
from those observed for EYM black holes in asymptotically flat \cite{89}
or de Sitter spacetime \cite{Torii:1995wv}.
Their regular counterparts are discussed in \cite{Bjoraker:2000qd, Bjoraker:2000yd}
and present also interesting properties.

In some sense, the minimal deviation from the asymptotically 
flat case is to include a "dual" or "magnetic" mass in the theory. 
That general relativity does admit configurations with "magnetic" masses
has been clear 
after the discovery of the NUT solution \cite{NUT,Misner} 
(see also \cite{Lynden-Bell:1996xj} for a recent discussion of this solution).
This metric has become renowned for being "a counterexample to
almost anything" \cite{misner-book} and represents a generalization of the 
Schwarzschild solution \cite{Hawking}. 
It is usually interpreted as describing a gravitational
dyon with both ordinary and magnetic mass.
It is the NUT charge which plays a dual role to ordinary  mass, 
in the same way that electric
and magnetic charges are dual within Maxwell theory \cite{dam}.
As discussed by many authors, 
the presence of magnetic-type mass (the NUT parameter $n$) 
introduces a "Dirac-string singularity" in the metric (but no curvature singularity).
This can be removed by appropriate identifications and changes in 
the topology of the spacetime manifold, which imply a periodic time coordinate. 
The Killing symmetries of this solution are still translations 
and $SO(3)$ rotations.
Spherical symmetry in a conventional sense is lost, though, 
when the NUT parameter is nonzero, since the 
rotations act on the time coordinate as well. 
The periodicity of the time coordinate prevents an interpretation 
of NUT metric as a black hole. 
Moreover, the metric is not asymptotically flat in the usual sense
although it does obey the required fall-off conditions.

An alternative suggestion due to Bonnor is to regard the string as a physical source \cite{Bonnor}.
In this approach, the source of the field is a spherically symmetric mass together with a 
semi-infinite massless source of angular momentum along the symmetry axis.

A large number of papers have been written 
investigating the properties 
of the gravitational analogs of magnetic monopoles.
The NUT solution has been generalized in a number of ways \cite{Bakler:cq}.
A charged solution of the Einstein-Maxwell equations with magnetic mass
has been found by Brill \cite{Brill}.
There are also nontrivial solutions of the low-energy  string theory
with NUT charge  (see e.g. \cite{Johnson:1994ek}).
In \cite{Ashtekar} a general framework was formulated to handle spacetimes with dual mass,
and showed that 
the dual mass is absolutely conserved in the presence of 
gravitational radiation.
Unfortunately, the pathology of closed timelike curves (CTC) is not 
special to the NUT solution but afflicts gravitation solutions with "dual" mass in general \cite{Magnon }.
The existence of magnetic mass requires either that the metric 
(not the curvature) be singular on a two-dimensional timelike world sheet 
or that the spacetime contains CTC.

As emphasized in \cite{Mueller:1986ij}, this condition emerges 
only from the asymptotic form of the fields,
and is completely insensitive to the precise details of the nature of the source, or the precise nature
of the theory of gravity at short distances where general relativity may be expected to break down.
The same construction can be made for a spacetime with a negative cosmological constant $\Lambda$, 
since it has also a well
defined spacelike infinity generated by a timelike Killing vector.
A similar analysis will yield the same pathologies.

Consequently,  spacetimes with NUT charge are not likely to be of interest as 
models for macroscopic objects.
Nevertheless, there are various features suggesting that their role 
cannot be neglected in 
the context of quantum gravity  \cite{Hawking:ig}.

A natural question arises: do there exist counterparts of the well-known hairy
black hole solutions possessing a magnetic mass?
And what new effects emerge due to the presence of a NUT charge?

Solutions of the YM equations on the Euclidean section of the Taub-NUT 
spacetime have been considered in the literature 
\cite{Pope:kj}, while a study of their Lorentzian counterparts is still  missing.
This fact is due perhaps to the widely belief that these spacetime are unphysical 
and thus cannot accomplish for interesting effects.

However, at the moment there are many indications that the NUT charge is an 
important ingredient in low energy string theory \cite{Johnson:1994ek}, 
conclusion enhanced by the discovery of "duality"
transformations which relate superficially very different configurations.
In many situations, if the NUT charge is not included in the study,
some symmetries of the system remain unnoticed (see e.g. \cite{Alonso-Alberca:2000cs}).
Therefore, we may expect to find a similar picture 
when discussing duality properties (yet to be discovered) of
 gravitating nonabelian fields.

There are also a number of reason to extend this analysis to configurations
with a negative cosmological constant.
Black holes with hair in such theories are useful for probing not only quantum 
gravity, but also may be interesting tests of AdS/CFT, particularly for classical
stable configurations.
Also, when $\Lambda<0$, the so-called topological black holes, whose topology
of the event horizon is no longer the two-sphere $S^2$ may appear
(see \cite{Mann:1997iz} for reviews of the subject).
These black holes embedded in 'locally AdS' background spacetimes (background locally
isometric to spacetimes of constant negative curvature) have been seminal 
to recent developments in black hole physics.
Vacuum Taub-NUT-AdS solution in four dimensions provided the first test bed for AdS/CFT 
correspondence in spacetimes where the asymptotic structure 
was only locally asymptotic AdS \cite{Hawking:1998ct}-\cite{Mann:1999pc}.
Their Euclideanized form have been a subject of interest in recent years,
in part because these asymptotically locally AdS spaces are relevant to 2+1
dimensional "exotic" CFT that live on the world volume of $M2$ branes 
after placing them on a squashed three sphere \cite{Dowker:1998pi}.
It is therefore natural to suppose that generalization
of these spacetimes for a nonabelian matter content might provide us with a better
understanding of these issues.
 
In this paper we restrict attention to $SU(2)$ EYM
theory  with $\Lambda \leq 0$,
and present numerical arguments for the existence of nontrivial solutions
with NUT charge. 
For a negative cosmological constant, these solutions present  nonvanishing nonabelian charges 
(with a suitable definition of charges in this case).
We find a family of solutions generalizing for a NUT charge the known spherically
symmetric dyon black hole solutions \cite{Bjoraker:2000qd}.
The Taub-NUT-AdS family of metrics contains also
solutions where the angular spheres $(\theta,\varphi$) are replaced by planes or
hyperboloids. 
For vanishing NUT charge, these solutions correspond to topological black holes.
In this paper we construct numerically the nonabelian dyon counterparts of these configurations,
generalizing for a NUT charge the topologically 
nonabelian black holes found in \cite{VanderBij:2001ia}.
The properties of these solutions do not differ too much from the properties of the
dyon black hole solutions (with $n=0$) in AdS spacetime.
In this case we find again a continuum of solutions possessing nonvanishing 
nonabelian charges and arbitrary mass.

However, the configurations with $\Lambda=0$ and NUT charge   
preserve the basic properties
of the well known hairy black hole solutions. 
The solutions do not posses nonabelian charges
(although a nonzero electric potential is always present) and, 
for a given $n$, the solution is uniquely determined by
the mass and the node number of the magnetic potential.

The  paper is structured as follows: in the next Section we explain the model 
and derive the basic equations and boundary conditions.
In Section 3  we show results obtained by numerical calculations 
and discuss the general properties of solutions as well as particular cases.
The boundary stress tensor
and the associated conserved charge are computed in Section 4, by using 
a counterterm prescription.
We discuss also the asymptotics of these spacetimes and find that a 
$(2+1)$-dimensional G\"odel-type universe is the boundary of a topologically NUT-AdS spacetime.
We conclude with Section 5 where the results are compiled.
\section{General framework and equations of motion}
\subsection{Metric ansatz and symmetries}
The EYM coupled system is described by the action
\begin{equation} 
\label{action}
I=\int d^{4}x\sqrt{-g}\left[\frac{1}{16\pi G} (R-2\Lambda)
-\frac{1}{4}F_{\mu \nu }^{a} F^{a\mu \nu }\right]
-\frac{1}{8\pi G}\int_{\partial\mathcal{M}} d^{3}x\sqrt{-h}K. 
\end{equation}
 
Here $G$ is the gravitational constant, $R$ is the Ricci scalar associated with the
spacetime metric $g_{\mu\nu}$ and 
$F_{\mu\nu}=\frac{1}{2} \tau^aF_{\mu\nu}^a$ is the gauge field strength tensor defined as
\begin{equation}
\label{fmn}
F_{\mu \nu} = 
\partial_\mu A_\nu -\partial_\nu A_\mu + i e \left[A_\mu , A_\nu \right] 
\ ,  
\end{equation}
where the gauge field is 
$ A_{\mu} = \frac{1}{2} \tau^a A_\mu^a,$
$\tau^a$ being the Pauli matrices. 
The last term in  (\ref{action}) is the Hawking-Gibbons surface term \cite{Gibbons:1976ue}, where $K$ is the trace 
of the extrinsic curvature for the boundary $\partial\mathcal{M}$ and $h$ is the induced 
metric of the boundary. Of course, this term does not affect the equations of motion 
but is important when discussing quantum properties of gravitating solutions.

Variation of the action (\ref{action})
 with respect to the metric $g^{\mu\nu}$ leads to the Einstein equations
\begin{equation}
\label{einstein-eqs}
R_{\mu\nu}-\frac{1}{2}g_{\mu\nu}R +\Lambda g_{\mu\nu}  = 8\pi G T_{\mu\nu},
\end{equation}
where the YM stress-energy tensor is 
\begin{eqnarray}
T_{\mu\nu} = 2{\rm Tr} 
    ( F_{\mu\alpha} F_{\nu\beta} g^{\alpha\beta}
   -\frac{1}{4} g_{\mu\nu} F_{\alpha\beta} F^{\alpha\beta}). 
\end{eqnarray}
Variation with respect to the gauge field $A_\mu$ 
leads to the matter field equations
\begin{equation}
\label{YM-eqs}
\nabla_{\mu}F^{\mu\nu}+ie[A_{\mu},F^{\mu\nu}]=0.
\end{equation}
We consider spacetimes whose metric can be written locally in the form
\begin{eqnarray}
\label{metric}
ds^2=\frac{dr^2}{N(r)}+R^2(r)(d\theta^{2}+f_k^{2}(\theta) d\varphi^{2})
-N(r)e^{-2\delta(r)}(dt+4 n f_k^2(\frac{\theta}{2}) d\varphi)^{2}.
\end{eqnarray}
The discrete parameter $k$ takes the values $1, 0$ and $-1$ 
and implies the form of the function $f_k(\theta)$
\begin{equation}
f_k(\theta)=\left \{
\begin{array}{ll}
\sin\theta, & {\rm for}\ \ k=1 \\
\theta , & {\rm for}\ \ k=0 \\
\sinh \theta, & {\rm for}\ \ k=-1.
\end{array} \right.
\end{equation}
We define the NUT parameter $n$, in terms of the coefficient appearing in the differential
$dt+4 n f_k^2(\theta/2)d\varphi$.

For $n=0$, metrics of this form have been considered by various authors when discussing
topological black holes with $\Lambda<0$. In this case,
$d \Omega_{(k)}^2=d\theta^{2}+f_k^{2}(\theta) d\varphi^{2}$
is the metric on a two-dimensional surface $\Sigma$ of constant curvature $2k$.
Thus, without a NUT charge, the topology of spacetime is $R^2\times H^2_g$, 
where $H^2_g$ is the topology of the surface $\Sigma$.
When $k=+1$, the universe takes on the familiar spherically symmetric form,
and the $(\theta, \varphi)$ sector has constant positive curvature. 
When $k=0$, the $\Sigma$ is a flat surface, while
for $k=-1$
the $(\theta, \varphi)$ sector is a space with constant negative curvature, 
also known as a hyperbolic plane. 
A discussion of these cases is available in \cite{Brill:1997mf}.
When $\Sigma$ is closed (which we'll suppose to be always the case), we denote its area by $V$,
defining $V=\int f_k (\theta) d \theta d \varphi$.

The form of $N(r),~\delta(r)$ and $R(r)$ emerges as result of demanding 
the metric to be a solution of the EYM equations with negative cosmological constant 
(note the existence of a metric gauge freedom which can be used to fix the expression of $R(r)$).

For this ansatz, the nonvanishing components of the Einstein tensor $E_{\mu}^{\nu}$ read
\begin{eqnarray}
\label{Eik}
\nonumber
E_r^r&=&-\frac{k}{R^2}-\frac{2N\delta'R'}{R}+\frac{NR'^2}{R^2}+\frac{N'R'}{R}
-e^{-2\delta}\frac{n^2N}{R^4},
\\
\nonumber
E_{\theta}^{\theta}&=&E_{\varphi}^{\varphi}=N\delta'^2-\frac{N\delta'R'}{R}-\frac{3\delta'N'}{2}+
\frac{N'R'}{R}-N\delta''+\frac{NR`'}{R}+\frac{1}{2}N''+e^{-2\delta}\frac{n^2N}{R^4},
\\
E_{t}^t&=&-\frac{k}{R^2}+\frac{2NR''}{R}+\frac{NR'^2}{R^2}+\frac{N'R'}{R}
-3e^{-2\delta}\frac{n^2N}{R^4},
\\
\nonumber
E_{\varphi}^t&=&4nf_k^2(\frac{\theta}{2})
\left(-\frac{k}{R^2}-N\delta'^2+\frac{N\delta'R'}{R}
+\frac{NR'^2}{R^2}+N\delta''+\frac{NR''}{R}-\frac{1}{2}N''+\frac{3\delta'N'}{2}
- 4e^{-2\delta}\frac{n^2N}{R^3}
\right),
\end{eqnarray}
where a prime denote a derivative with respect to $r$.

Apart from the Killing vector $\partial_{t}$, this line element possesses 
three more Killing vectors
characterizing the NUT symmetries
\begin{eqnarray}
\label{Killing}
\nonumber
K_1&=&\sin \varphi \partial_{\theta}
+\cos \varphi \frac{f_k'(\theta)}{f_k(\theta)}\partial_{\varphi}
+2n\cos\varphi \frac{f_k(\theta/2)}{f_k'(\theta/2)}\partial_{t},
\\
K_2&=&\cos \varphi \partial_{\theta}
-\sin \varphi \frac{f_k'(\theta)}{f_k(\theta)}\partial_{\varphi}
-2n\sin \varphi \frac{f_k(\theta/2)}{f_k'(\theta/2)}\partial_{t},
\\
\nonumber
K_3&=&\partial_{\varphi}-2n\partial_{t}.
\end{eqnarray}
These  Killing vectors form a subgroup with the same structure constants 
that are obeyed by spherically symmetric
solutions $[K_a,K_b]=\epsilon_{abc}K_c$.

The properties of case $k=1$ are well known from the study of Taub-NUT solution.
The rotational symmetries act in the usual way on the angular coordinates,
but also involve time translations in order to preserve
the differential $dt+4n\sin^2 (\theta/2) d \phi$. 
Thus we have lost spherical symmetry, at least in the conventional sense. 

Again, the $n\sin^2 (\theta/2)$ term in the metric means that a small loop around the 
$z-$axis does not shrink to zero at $\theta=0$ and at $\theta=\pi$.
These singularities can be regarded ar the analogue of a Dirac string in electrodynamics
and  are not the usual degeneracies of spherical coordinates on the two-sphere.
This problem was first encountered in the vacuum NUT metric.
One way to deal with this singularity has been proposed by Misner \cite{misner-book}.
His argument holds also independently of the precise functional form of $N$ and $\delta$.
In his construction, one considers one coordinate patch in which the string runs off to
infinity along the north axis.
A new coordinate system can then be 
found with the string running off to infinity along the south axis.
This indicates that the string is an artifact resulting from a poor choice of coordinates.
In the intersection of these two coordinate patches, a necessary condition for consistency is to
identify the time coordinate with period $8 \pi n$.
The coordinate singularity at $\theta=\pi$ can be removed via the introduction
of 
\begin{eqnarray}
\label{transf}
t'=t+4n\varphi. 
\end{eqnarray}  
It is clear that the $t$ coordinate is also periodic with period 
$8 \pi n$ and essentially becomes an Euler angle coordinate on $S^3$.
Thus there are CTC  which can not be removed by going to a covering space 
and no reasonable spacelike surface.
One finds also that surfaces of constant radius have the topology
of a three-sphere, in which there is a Hopf fibration of the $S^1$
of time over the spatial $S^2$ \cite{misner-book}.

Therefore for $n$ different from zero, the $k=1$ metric structure (\ref{metric}) 
shares the same troubles 
exhibited by the Taub-NUT gravitational field in the presence, if any, 
of an electromagnetic field, $i.e.$
the solutions cannot be interpreted properly as black hole.
However, for $k=0,-1$, the fibration is trivial and there are no Misner strings \cite{Emparan:1999pm},
but again, there are no reasonable spacelike surfaces.

One can remark another interesting property of the line element (\ref{metric}).
The static nature of a hairy spherically symmetric  black hole solution implies that 
it can only produce a "gravitoelectric" field.
In the case of a NUT charge, the existence of the cross metric term $d\varphi dt$ 
shows that this space has also a "gravomagnetic" field.
Thus the line-element (\ref{metric}) it is Kerr-like in the sense that it has a crossed 
 metric component $g_{\varphi t}$
which generates "gravomagnetic" effects. 
This term does not produce an ergoregion but it does produce 
an effect similar to the dragging of inertial
frames \cite{Zimmerman:kv}.

\subsection{$SU(2)$ connection}
The most general expression for the appropriate $SU(2)$ connection 
is obtained by using the standard rule 
for calculating the gauge potentials for any spacetime group 
\cite{Forgacs:1980zs,Bergmann:fi}.
According to Forgacs and Manton, a gauge field admit a spacetime symmetry if the spacetime
transformation of the potential can be compensated 
by a gauge transformation \cite{Forgacs:1980zs}.

Taking into account the symmetries of the line element (\ref{metric})  
we find the general form
\begin{eqnarray}
\label{A}
A=\frac{1}{2e} \left \{  \Big(dt+4n f_k^2(\frac{\theta}{2}) d\varphi\Big)u(r) \tau_3+
\nu(r) \tau_3 dr+
\Big(\omega(r) \tau_1 +\tilde{\omega}(r) \tau_2\Big) d \theta
+ \Big[f_k'(\theta) \tau_3+
(\omega(r) \tau_2-\tilde{\omega}(r)\tau_1  ) f_k(\theta) \Big]d \varphi  \right\},
\end{eqnarray}
(see Ref. \cite{Bjoraker:2000qd}
for an extensive discussion of this ansatz in the case $k=1,~n=0$).
The gauge connection (\ref{A}) remains invariant under a residual $U(1)$ gauge symmetry 
which can be used to set
$\nu=0$. Because the variables $\omega$ and $\tilde{\omega}$ 
appear completely symmetrically in the EYM system, the two amplitudes 
must be proportional and we can always set
$\tilde{\omega}=0$ (after a suitable gauge transformation).
By using the residual $U(1)$ gauge symmetry, we can also consistently set
$\nu= 0$. 
Hence, without any loss of generality, the gauge potential is described by two 
functions $\omega(r)$ and $u(r)$ which we shall refer to as magnetic 
and electric potential, respectively.

Therefore, the components of the  field strength tensor are
\begin{eqnarray}
\label{field}
F_{r\theta}&=&\frac{\tau_1}{2}\omega',~~~
F_{r \varphi}=\frac{\tau_2}{2}\omega' f_k(\theta)
+\frac{\tau_3}{2} 4n u' f_k^2(\frac{\theta}{2}),~~~
F_{r t}=\frac{\tau_3}{2}u',
\\
\nonumber
F_{\theta \varphi}&=&-\frac{\tau_2}{2} 4nu\omega f_k^2(\frac{\theta}{2}) +
\frac{\tau_3}{2}(\omega^2-k+2nu)f_k(\theta),~~~
F_{\theta t}=-\frac{\tau_2}{2}u\omega,~~~
F_{\varphi t}=\frac{\tau_1}{2}u \omega f_k(\theta). 
\end{eqnarray}
It is difficult to find a satisfactory  definition of the 
nonabelian electric and magnetic charges in the presence
of a NUT parameter.
For $n=0$, the usual (-gauge dependent) definition is
\begin{equation}
\label{def-charge}
\pmatrix{Q_E\cr Q_M\cr}
 = {e\over 4\pi} \int dS^{(k)}_{\mu} \, \sqrt{-g} \,
Tr\Big\{ 
 \pmatrix{ F^{\mu t}\cr \tilde F^{\mu t} \cr} \tau_3 
 \Big\},
\end{equation}
where  $\tilde F^{\mu \nu}$ is the dual field strength tensor.
However, the integral (\ref{def-charge}) does not intrinsically characterize a field 
configuration, as it is gauge dependent \cite{Bjoraker:2000qd}.
The definition of conserved currents and charges in a non-Abelian Yang-Mills theory 
is a problem approached  by different authors in the last decades 
(see $e.g.$ \cite{Chrusciel:1987jr}-\cite{Creighton:1995au}), with various solutions.
A gauge invariant definition for the nonabelian charges has been proposed in  
\cite{Corichi:1999nw}  (see also \cite{Kleihaus:2002ee})
\begin{eqnarray}
\label{charges}
Q_E={e\over 4\pi} \oint d \theta d \varphi |\tilde F_{\theta \varphi}|,~~~
Q_M={e\over 4\pi} \oint d \theta d \varphi | F_{\theta \varphi}|,
\end{eqnarray}
where the vertical bars denote the Lie-algebra norm and the integrals are evaluated as $r \to \infty$.
One can verify that this definition yields the absolut value 
of the charges as defined in (\ref{def-charge}).
Also, for the field configuration we consider here,
the same expression (\ref{def-charge}) for $Q_E,~Q_M$ is obtained when making use 
of the existence of a Lie-algebra valued "Killing scalar"
\cite{Volkov:1999cc, Creighton:1995au}.

In the presence of a dual mass ($n \neq 0$) it seems that we can only define 
the magnetic and electric charges 
in terms of the asymptotic behavior of the gauge field 
by analogy to that in asymptotically flat spacetime, 
$i.e.$ from (\ref{field}) 
$F_{tr}^{(3)} \simeq Q_E/r^2$ and $F_{\theta \phi}^{(3)} \simeq Q_M f_k(\theta) $ 
(a similar problem occurs for an $U(1)$ field
\cite{Johnson:1994ek}).

We now remove the dependence on the coupling 
constants $G$ and $e$ from the differential
equations by using the following rescaling $r \to r \sqrt{4 \pi G}/e,~n \to n \sqrt{4 \pi G}/e,~
u \to u e/\sqrt{4 \pi G},~\Lambda \to \Lambda e^2/(4 \pi G)$.

For this ansatz, the nonvanishing components of the energy-momentum tensor read
\begin{eqnarray}
\label{Tik}
\nonumber
T_r^r&=&\frac{N \omega'^2}{R^2}-\frac{1}{2}u'^2 e^{2 \delta}
-\frac{(\omega^2-k+2nu)^2}{2R^4}+ e^{2 \delta}\frac{\omega^2 u^2}{R^2N},
\\
\nonumber
T_{\theta}^{\theta}&=&T_{\varphi}^{\varphi}= 
\frac{1}{2}u'^2 e^{2 \delta}+\frac{(\omega^2-k+2nu)^2}{2R^4},
\\
T_t^t&=&-\frac{N \omega'^2}{R^2}-\frac{1}{2}u'^2 e^{2 \delta}
-\frac{(\omega^2-k+2nu)^2}{2R^4}-e^{2 \delta}\frac{ \omega^2 u^2}{R^2N},
\\
\nonumber
T_{\varphi}^t&=&-4nf_k^2\big(\frac{\theta}{2}\big)\left(
\frac{N \omega'^2}{R^2}+u'^2 e^{2 \delta}
+\frac{(\omega^2-k+2nu)^2}{ R^4}+e^{2 \delta}\frac{ \omega^2 u^2}{R^2N}
\right).
\end{eqnarray}
\subsection{Field equations}
One explicit solution of the EYM equations
is well known. This is the charged generalization of the NUT 
solution found by Brill \cite{Brill}, generalized for a nonvanishing $\Lambda$ \cite{kramer}.
These abelian configurations have a
mass $M$, electric charge $Q$ and magnetic charge $P=k$
\begin{eqnarray}
\label{Brill}
\omega&=&0,~~~~~u=\frac{nP-Qr}{r^2+n^2},~~~~\delta=0,
\nonumber
\\
R(r)&=&\sqrt{r^2+n^2},~~~
N=k-\frac{2(Mr+kn^2)+(Q^2+P^2)}{r^2+n^2}-\frac{\Lambda}{3}\frac{r^4+6r^2 n^2-3n^4}{r^2+n^2}.
\end{eqnarray}
Some features of these Einstein-Maxwell solutions 
are discussed in \cite{Alonso-Alberca:2000cs}.
The properties of the vacuum topologically-Taub-NUT-AdS solutions (on 
the Euclidean section) are discussed in \cite{Emparan:1999pm}.

The Brill solution is written in a metric gauge $\delta=0$. 
For a nonabelian configuration, a more convenient gauge, used also in 
the numerical computation is the Schwarzschild one
$R(r)=r$.
For this choice, we find from (\ref{YM-eqs}) that the nonabelian potentials 
should satisfy the equations
\begin{eqnarray}
\label{eq1}
(N e^{-\delta}\omega')'&=&\frac{ e^{-\delta}\omega(\omega^2-k+2nu)}{r^2}
-\frac{e^{\delta} \omega u^2}{N},
\\
\nonumber
(e^{\delta}u'r^2)'&=&\frac{ 2e^{\delta}\omega^2 u}{N}
-\frac{2n e^{-\delta}(\omega^2-k+2nu)}{r^2},
\end{eqnarray}
while the Einstein equations (\ref{einstein-eqs}) imply
\begin{eqnarray}
\label{eq3}
(e^{-\delta})'+\frac{n^2}{r^3}e^{-3\delta}-
\frac{2}{r}\left(e^{-\delta}\omega'^2+e^{\delta}\frac{u^2 \omega^2}{N^2}\right)=0,
\\
\nonumber
\label{eq4}
rN'-k+N+\Lambda r^2-\frac{3n^2N}{r^2}e^{-2\delta}+2N\omega'^2+
\frac{(\omega^2-k+2nu)^2}{r^2}+2e^{2\delta}
\left(\frac{r^2u'^2}{2}+\frac{u^2\omega^2}{N}\right)=0.
\end{eqnarray}
Note that, for $n \neq 0$, these equations do not present the usual symmetry $u \to -u$, 
which is related to the absence of time reflection symmetry in the metric (\ref{metric}).
Also, for an asymptotically flat hairy black hole, 
the vanishing of the electric potential
is imposed by finite energy requirements \cite{bizon}.
This conclusion remain valid in the presence of a positive cosmological constant \cite{Bjoraker:2000qd}.
However, for $n \neq 0$ we cannot consistently set $u=0$ for nonabelian solutions.
Thus, all solutions with NUT charge  necessarily present 
an electric part and possibly an electric charge.

\subsection{Boundary conditions}
We want the metric (\ref{metric}) to describe a nonsingular, asymptotically 
NUT spacetime outside an
horizon located at $r=r_h$. 
Here $N(r_h)=0$ is only a coordinate singularity where all curvature invariants are finite.
A nonsingular extension across 
this null surface can be found just as at the event horizon of a black hole.
If the time is chosen to be periodic, 
as discussed above, this surface would not be a 
global event horizon, although it would still
be an apparent horizon.
The parameter $n$ describes the departure of the metric (\ref{metric}) 
from the static black hole solutions.
The regularity assumption implies that all curvature invariants at $r=r_h$
are finite.
Also, the energy density $T_t^t$ is assumed to be regular everywhere.

We find that the general properties and asymptotic form of the solutions crucially 
depend on the presence of $\Lambda$.
\subsubsection{The case $\Lambda<0$}
For $\Lambda<0$, we find the following asymptotic form of the solution
\begin{eqnarray}
\label{i1}
\nonumber
N &\sim& k-n^2\Lambda-\frac{2M}{r}-\frac{\Lambda}{3}r^2+O\left(\frac{1}{r^2}\right),
\\
\nonumber
e^{-\delta} &\sim& 1+\frac{n^2}{2r^2}+ O\left(\frac{1}{r^4}\right),
\\
\omega &\sim& \omega_0+\frac{\tilde{\omega}_1}{r}+\frac{\omega_2}{r^2}+O\left(\frac{1}{r^3}\right),
\\
\nonumber
u &\sim& u_0+\frac{\tilde{u}_1}{r}+\frac{\tilde{u}_2}{r^2}+O\left(\frac{1}{r^3}\right),
\end{eqnarray}
where 
\begin{eqnarray}
\omega_2&=&-\frac{3\omega_0}{2\Lambda}(\omega_0^2-k+2n u_0)
-\frac{9\omega_0u_0^2}{2\Lambda^2},
\\
\nonumber
\tilde{u}_2&=&-\frac{3\omega_0^2u_0}{\Lambda}-n(\omega_0^2-k+2nu_0)^2,
\end{eqnarray}
and $\omega_0,~u_0,~\tilde{u}_1,~\tilde{\omega}_1$ are constants to be determined. 
Thus,  the geometry approaches asymptotically the (topological-) Taub-NUT-AdS spacetime.

For these asymptotics we define the nonabelian charges  
\begin{eqnarray}
\pmatrix{Q_E\cr Q_M\cr}
 = \frac{V }{4 \pi} \pmatrix{ \tilde{u}_1 \cr k - w_0^2-2nu_0\cr}.
\end{eqnarray}
In the rest of this paper $Q_E,~Q_M$ are expressed in units of $V/4 \pi$. 

The corresponding expansion as $r \to r_h$ is
\begin{eqnarray}
\label{eh}
\nonumber
N(r)&=&N_h(r-r_h)+O\left((r-r_h)^2\right),
\\
\nonumber
e^{-\delta(r)}&=&e^{-d_0}+d_1(r-r_h)+O\left((r-r_h)^2\right),
\\
\omega(r)&=&\omega_h+\omega_1(r-r_h)+O\left((r-r_h)^2\right),
\\
\nonumber
u(r)&=&u_1(r-r_h)+u_2(r-r_h)^2+O\left((r-r_h)^3\right),
\end{eqnarray}
where 
\begin{eqnarray}
\nonumber
N_h&=&\frac{1}{r_h}\left(k-\Lambda r_h^2-\frac{(\omega_h^2-k)^2}{r_h^2}-e^{2d_0}u_1^2 r_h^2\right),
\\
d_1&=&\frac{2}{r_h}e^{-d_0}\left(\omega_1^2-\frac{n^2}{2r_h^2}+
e^{2d_0}\frac{\omega_h^2 u_1^2}{N_h^2}\right),
\\
\nonumber
\omega_1&=&\frac{\omega_h(\omega_h^2-k)}{r_h^2N_h},
\\
\nonumber
u_2&=&\frac{\omega_h^2 u_1}{N_h r_h^2}-e^{-2d_0}\frac{n (\omega^2_h-k)}
{r_h^4}+e^{-d_0}\frac{u_1d_1}{2}.
\end{eqnarray}
In the numerical computation we have set $d_0=1$, without loss of generality.
The usual condition $\delta(\infty)=0$ is connected with our choice by
a a simple time-coordinate rescaling, 
together with a rescaling of $u$ and $n$ 
(we observe that the field equations (\ref{eq1})-(\ref{eq3}) remain unchanged if we take
$t \to t/c,~\delta \to \delta-\log c,~u \to cu , n \to n/c$).
The magnetic charge is not affected by this rescaling, 
while for the electric charge we have
$Q_E \to c Q_E $.

The function $N$ should also satisfy the obvious condition
$ 
\frac {dN(r)}{dr}
\Big| _{r_{h}} >0$. 
For given ($r_h,~\Lambda$), this places a bound on $\omega_h$
\begin{eqnarray}
\label{mhbound}
\frac {\left( \omega_h^{2}-k \right) ^{2}}{r_h ^{2}}+u_1^2r_h^2
<k-\Lambda r_h ^{2},
\end{eqnarray}
and implies the positiveness of the quantity $\omega'(r_h)$.
In the $k=-1$ case, this implies the existence of 
a minimal value of $|\Lambda|$, $i.e.$ for a given $r_h$
\begin{eqnarray}
|\Lambda|>\frac{1}{r_h^2}(1+\frac{1}{r_h^2}).
\end{eqnarray}
The asymptotic behavior (\ref{i1}) suggest to adopt the following 
form for the metric function $N(r)$
\begin{eqnarray}
\label{ans}
N(r)=k-n^2\Lambda-\frac{2 m(r)}{r}-\frac{\Lambda r^2}{3}.
\end{eqnarray}
For $n=0$, $m(r)$ corresponds to the local mass-energy function and 
its limit as $r\to \infty$ corresponds
to ADM mass.
We present arguments in Section 4 that this interpretation 
is still valid for a nonvanishing NUT charge.
The numerical results have been obtained for this form of $N(r)$ 
which factorizes the divergent part
$\Lambda r^2/3$.
\subsubsection{The case $\Lambda=0$}
As expected, for $\Lambda=0$, the behavior of the solutions at infinity is very different
as compared to the case $\Lambda<0$.
The only allowed configuration for a vanishing cosmological constant is $k=1$.
In this case, we find the asymptotic form of the solutions
\begin{eqnarray}
\label{asf}
\nonumber
N&\sim&1-\frac{2M}{r}-\frac{3n^2}{r^2}+\frac{3n^2M}{r^3}+
\frac{2(\tilde{\omega}_1^2+\tilde{u}_2^2+n^4)}{r^4}+O\left(\frac{1}{r^5}\right),
\\
\nonumber
e^{-\delta}&\sim&1+\frac{n^2}{2r^2}+\frac{d_4}{r^4}+O\left(\frac{1}{r^5}\right),
\\
\omega&\sim&\pm1+\frac{\tilde{\omega}_1}{r}+O\left(\frac{1}{r^2}\right),
\\
\nonumber
u&\sim&\frac{\tilde{u}_2}{r^2}+O\left(\frac{1}{r^3}\right),
\end{eqnarray}
where $\tilde{\omega}_1,~\tilde{u}_2,~d_4$ are constants to be determined.
Again $M$ is to be interpreted as the total mass of the solutions.
The boundary conditions on the event horizon are obtained from
(\ref{eh}) by taking the limit $\Lambda=0$.

In asymptotically Minkowski spacetime ($n=0$), 
the electric components of the YM connection are forbidden in static solutions \cite{bizon}.
We find that this is not valid for solutions possessing a NUT parameter.
Moreover, as seen from the YM equations, the existence of a nonzero electric potential is
required in this case.
This has to be associated with the Kerr-like form of the line-element (\ref{metric}).
A similar property is shared by axially symmetric rotating black 
holes in asymptotically flat spacetime
\cite{Kleihaus:2002ee}.
 
We remark also that, for $n\neq 0$, the decay of $u$ is too rapid to give 
a net electric charge and $\omega(\infty)^2=1$ corresponds to zero 
magnetic charge.
Thus, these nonabelian solutions are not dyons in the usual sense but do possess both
electric and magnetic fields.
A similar property has been noticed for the asymptotically 
flat solutions of an EYM-massive axion theory \cite{O'Neill:1993qz}.

\section{Numerical results}
Although an analytic or approximate solution still appears to be intractable,
we present in this section numerical arguments that the known EYM black hole solutions
can be extended to include a NUT parameter.
The properties of the solutions in the cases $\Lambda<0$ and $\Lambda=0$ are very different and
we will discuss them separately.

\subsection{Charged topological black holes}
We start by discussing the particular case $n=0,\Lambda<0$.
The properties of the spherically symmetric dyon solutions $k=1$ are studied in 
Ref. \cite{Bjoraker:2000qd} (see also \cite{Radu:2001ij}).
In the presence of a negative cosmological constant, 
the "nonabelian baldness" theorems (forbidding $SU(2)$ dyons) are no longer valid: 
there are essentially nonabelian solutions with an electric sector.
Dyon solutions of the EYM equations are determined if the parameter $u_1$ in (\ref{eh}) is chosen 
to be nonzero for a given $\Lambda$.
One finds dyon solutions with finite mass for continuous varying $\omega _{h}$ 
(and the corresponding parameter in the electric sector) and any number of 
nodes $p$ including $p=0$.

We shall therefore focus on the cases $k=0$ and $k=-1$.
Although predicted in \cite{VanderBij:2001ia},  
these topological dyon black holes are not discussed in literature.
An analysis of the 
properties of corresponding monopole solutions  
is presented in \cite{VanderBij:2001ia}.
These are black hole solutions with nonspherical event horizon topology and possessing a 
nonvanishing magnetic potential $\omega(r)$.
For $u=0$  only nodeless solutions are allowed.
Also all $k=0$ and the $k=-1$ solutions with $|\omega_0|>1$ are found to be stable 
against linear fluctuations.
Another interesting property is the existence of $k=-1$ solutions with $M<0$, a common property,
however, of this class of topological black holes \cite{Brill:1997mf}.
Black hole solutions with nonspherical topology do not have globally regular 
counterparts.

For $n=0$ and $k=0,-1$, we numerically solve the full set of Eqs. (\ref{eq1})-(\ref{eq4})
with the boundary conditions (\ref{i1}) 
using a standard shooting method. 
Starting with the horizon expansion
(\ref{eh}), the equations were integrated 
for a range of values of $(r_h,~\Lambda)$ 
and varying $(\omega _{h},~u_1$). 
Since in this case the field equations are invariant 
under the transformation $\omega \to -\omega,~u\to -u$, 
only values of $\omega_h,~u_1$ greater than zero are considered.
For sufficiently small $\omega_h$, all field variables remain 
close to their values for the abelian configuration
 with the same $r_h$.
Significant differences occur for large enough values of $\omega_h$
and the effect of the nonabelian field on the geometry 
becomes more and more pronounced.

The solutions we obtain have many common properties with their
spherically symmetric $\Lambda<0$ counterparts.
Solutions are classified by the parameters $M$, $Q_M$ and $Q_E$.
The behavior of metric functions $m$ and $\delta$ is similar to the spherically symmetric 
monopole black hole solutions.
For every considered value of $(\Lambda,~r_h)$, we find regular black hole solutions
for  compact intervals of $\omega_h$ and $u_1$. 
For $\omega_h>\omega_h^c$ and $u_1>u_1^c$ the solutions blow up 
or the function $N(r)$ becomes negative.
The values of $\omega_h^c,~u_1^c$ increase as $|\Lambda |$ 
increases and depend also on the value of $r_h$.
There are also solutions for which $\omega_{\infty}>1$
although $\omega_h<1$.

In contrast to the monopole case, 
both nodeless solutions and solutions where $\omega$ crosses the axis can exist.
This is not a surprise, since the simple argument presented in \cite{VanderBij:2001ia}
for the existence of nodeless solutions only is invalidated by the presence of an electric potential. 
For $k=0$ we have always $m(r)>0$ and the black holes 
therefore only occur with positive values
of mass. 
 Typical solutions in these cases are presented in Figure 1.
In Figure 2 we present the mass parameter $M$, the nonabelian charges $Q_{E},Q_M$
and the values of the gauge potentials
at infinity $u_0,\omega_0$ as a function of the magnetic potential value 
on the horizon for fixed $r_h,~\Lambda$.
Not unexpectedly, when the surface $\Sigma$ has a negative curvature,
we have found for suitable values of ($\Lambda,~r_h,~\omega_h,~u_1$),
a class of solutions with $M<0$ or even $M=0$, 
apart from configurations with positive $M$.

\subsection{Solutions with NUT charge and negative cosmological constant}
Our approach for a nonvanishing $n$ is very similar to the black hole case.
By using a standard shooting method, we integrate the 
field equations (\ref{eq1})-(\ref{eq4}) in the general case.
The NUT charge $n$, as well as the cosmological constant $\Lambda$ are primary input parameters.
We  call $r_h, ~\omega _{h},~u_1$ 
secondary input parameters.
A complete analysis of the complex correlation between the  parameters
of the theory  is beyond the purposes of this paper. Here we present only the results of a
preliminary investigation.
However, we found nontrivial solutions for a large range of primary parameters $n,\Lambda$, 
some fixed values of $r_h$
and compact intervals of secondary parameters $\omega _{h},~u_1$.
Nontrivial solutions seems to exist for every choice of $n,\Lambda $ satisfying (\ref{mhbound}).

The numerically solutions we find in this case 
retains all the features of 
the "gravitoelectric" case $n=0$. 
Also, the fundamental properties of the solutions do not depend on the value of $k$.
For small values of the NUT parameter the solutions do not considerably deviate from the 
corresponding black hole solutions.

For fixed primary input parameters and a given $r_h$, solutions are found for  continuous
sets of secondary parameters $\omega _{h}$ and $u_1$.
Again we find a continuum of solutions where $w$ crosses the axis 
an arbitrary number of times depending on $\omega _{h}$ and $u_1$.
The existence of critical secondary parameters $\omega _{h}^c$ and $u_1^c$ is also noticed.
Similarly there exist solutions where $\omega$ does not cross the axis. 
For $k=-1$ we find a class of solutions with $M<0$ or even $M=0$, 
apart from configurations with positive $M$.
Typical solutions are presented in Figure 3. 
We can see that the electric potential starts at zero 
and asymptotically approaches a finite value $u_0$. 
The mass parameter stay also finite.
In Figure 4 we present the results of the numerical integration for a varying  NUT  charge, while keeping
$(\Lambda,~r_h,\omega_h,~u_1)$ fixed.
In Figure 5 we vary the secondary parameter $u_1$ for fixed $(\Lambda,~n,~r_h,\omega_h)$.
Solutions with nonavanishing nonabelian charges are found also for $u_1=0$, $\omega_h\neq0$.

\subsection{Solutions with NUT charge and $\Lambda=0$}
The case $\Lambda=0,~k=1$ is rather special.
Such a spacetime is not asymptotically flat in the usual sense, 
although it does obey the required fall-off
conditions. 

The solutions we find are  a 
kind of extension of the well known asymptotically flat black hole solutions \cite{89}.
As a new feature, they present a nonvanishing electric potential but no nonabelian charges.

We find that, given ($n, r_h$), 
solutions may exist for a discrete set of shooting parameters
 $\omega_0$ and $u_1$.
By using a standard ordinary  differential  
equation solver, we
evaluate  the  initial  conditions  at  $r_h$ for  global  tolerance  
$10^{-12}$, adjusting  for fixed shooting parameters and  integrating  towards  $r\to\infty$.
The integration stops when the asymptotic
limit (\ref{asf}) is reached.

The profile of typical solutions is presented in Figure 6.
We observe that the electric potential starts at zero and monotonically increases rapidly
to some maximal value, afterward decreasing slowly to zero.
The behavior of the $\omega,~m,~\delta$ in the region
$r>r_h$ is qualitatively similar to that for the black hole solutions.
The function $\omega$ starts from some value $0<\omega_h<1$ at the horizon and after
$p$ oscillations around zero tends asymptotically to $(-1)^p$.
We have found solutions up to $p=3$, 
but we have no reason to doubt 
the existence of similar solutions for every value of $p$.
For a given $n$, the geometry in the asymptotic region $r \to \infty$ is Taub-NUT, 
with a mass $M=M_p(r_h)$.
Similar to the black hole case, the node number $p$ cannot be associated with any kind
of YM charge.

For all solutions, we found that always $|\omega|<1$ everywhere outside the horizon.
We noticed that, for a given $p$, the value of the mass-parameter $M$ for $n \neq 0$,
is always smaller that the corresponding black hole  ADM mass.
For example, the known value for $n=0$ is $M=0.937$; for $n=0.284$ we found $M=0.887$ while
for $n=0.493$, $M=0.754$ (this is valid for $p=1,~r_h=1$).

Another point which deserves further research is the existence in this case 
of a maximal value of $n$.
For a given value of $r_h$ we failed to find numerical solutions with the right asymptotics 
beyond a value $n_{max}$ of the NUT parameter.
Since we deal with a two-dimensional  shooting problem,
it is difficult to find  accurate values for the $n_{max}$.
For example, for configurations with $r_h=1,~p=2$, we find no regular solutions beyond $n>0.21$.

\section{A computation of the stress tensor and the mass}

In order to compute quantities like the ADM mass one usually encounters
infrared divergences, which are regularized
by subtracting a suitable chosen background
\cite{Gibbons:1976ue,Hawking:ig}. 
Such a procedure, however, in general is not unique; 
in some
cases (such as solutions with NUT charge) the choice of reference background is ambiguous.
Recently, in order to regularize such divergences, a different
procedure has been proposed \cite{Balasubramanian:1999re}.
This technique was inspired by AdS/CFT correspondence and consists in adding suitable counterterms 
$I_{ct}$
to the action. These counterterms are built up with
curvature invariants of a boundary $\partial \cal{M}$ (which is sent to 
infinity after the integration)
and thus obviously they do not alter the bulk equations of motion.
The following counterterms are sufficient to cancel divergences in four dimensions,
for vacuum solutions with a negative cosmological constant
\begin{eqnarray}
\label{ct}
I_{\rm ct}=-\frac{1}{8 \pi} \int_{\partial {\cal M}}d^{3}x\sqrt{-h}\Biggl[
\frac{2}{l}+\frac{l}{2}\cal{R}
\Bigg]\ .
\end{eqnarray}
Here ${\cal R}$ is the Ricci scalar for the boundary metric $h$. 
In this section we will define also $l^2=-3/\Lambda$. 

Using these counterterms one can
construct a divergence-free stress tensor from the total action
$I{=}I_{\rm bulk}{+}I_{\rm surf}{+}I_{\rm ct}$ by defining 
\begin{eqnarray}
\label{s1}
T_{ab}&=& \frac{2}{\sqrt{-h}} \frac{\delta I}{ \delta h^{ab}}
=\frac{1}{8\pi }(K_{ab}-Kh_{ab}-\frac{2}{l}h_{ab}+lE_{ab}),
\end{eqnarray}
where $E_{ab}$ is the Einstein tensor of the intrinsic metric $h_{ab}$.
The efficiency of this approach has been demonstrated in a broad range of examples.
The counterterm subtraction method has been developed on its own interest and applications.
If there are matter fields on $\cal{M}$ additional counterterms may be needed to regulate the action.
However, we find that for a $SU(2)$ nonabelian matter content 
the prescription  (\ref{ct}) removes all divergences.
The results we find by using the asymptotic expressions (\ref{i1}) 
for the boundary stress tensor at large $r$ 
are 
\begin{eqnarray}
\label{BD4}
\nonumber
T_{\theta \theta} &=&\frac{1}{8 \pi}\frac{lM}{r}+O\left(\frac{1}{r^2}\right),
\\
T_{\varphi \varphi}&=&\frac{1}{8 \pi}\left(\frac{32Mn^2}{l}f_k^4(\frac{\theta}{2})
+lM f_k^2(\theta)\right)\frac{1}{r}+O\left(\frac{1}{r^2}\right),
\\
\nonumber
T_{\varphi t}&=&\frac{1}{8 \pi}8Mnf_k^2(\frac{\theta}{2})\frac{1}{lr}+O\left(\frac{1}{r^2}\right),
\\
\nonumber
T_{tt}&=&\frac{1}{8 \pi}\frac{2M}{lr}+O\left(\frac{1}{r^2}\right).
\end{eqnarray}
Direct computation shows that this stress tensor is traceless.
This result is expected from the AdS/CFT correspondence, since even dimensional
bulk theories with $\Lambda<0$ are dual to odd dimensional CFTs which have a 
vanishing trace anomaly.

We can use this formalism to assign a mass to our EYM solutions
by writing the boundary metric in an ADM form \cite{Balasubramanian:1999re}
\begin{eqnarray}
\label{b-AdS}
h_{\mu \nu}dx^{\mu} dx^{\nu}=-N_{\Sigma}^2dt^2
+\sigma_{ab}(dx^a+N_{\sigma}^a dt) (dx^b+N_{\sigma}^b dt).
\end{eqnarray}
If $\xi^{\mu}$ is a Killing vector generating an isometry of the boundary geometry,
there should be an associated conserved charge. The conserved charge associated with
time translation is the mass of spacetime
\begin{eqnarray}
\label{mass}
\mathbf{M}=\int_{\partial \Sigma}d^{D-1}x\sqrt{\sigma}N_{\Sigma}\epsilon.
\end{eqnarray}
Here $\epsilon=n^{\mu}n^{\nu}T_{\mu \nu}$ is the proper energy density while $n^{\mu}$ 
is a timelike unit normal to $\Sigma$.
By using this relation, we can find the mass of our solutions, 
which is always $MV/4 \pi$, as expected.
This result keeps valid in the limit $k=1,\Lambda=0$.

The metric restricted to the boundary $h_{ab}$ diverges due to an infinite
conformal factor $r^2/l^2$. The background metric upon which the dual field
theory resides is
\begin{eqnarray}
\gamma_{ab}=\lim_{r \rightarrow \infty} \frac{l^2}{r^2}h_{ab}.
\end{eqnarray}
For the asymptotically topologically-Taub-NUT-AdS solutions considered here, the  
boundary metric is 
\begin{equation}
\label{l1} 
\gamma_{ab}dx^a dx^b=l^2(d \theta^2+f_k^2(\theta)d \varphi^2)-
(dt+4nf_k^2(\frac{\theta}{2})d\varphi)^2.
\end{equation}
In the light of AdS/CFT correspondence, Balasumbramanian and Kraus have interpreted 
Eq. (\ref{s1}) as $\tau^{ab}=\frac{2}{\sqrt{-\gamma}} \frac{\delta S_{eff}}{\delta \gamma_{ab}}$
where $\tau^{ab}$ is the expectation value of the CFT stress tensor.
Corresponding to the boundary metric (\ref{l1}), the stress-energy tensor $\tau_{ab}$
for the dual theory can be calculated using the following relation \cite{Myers:1999qn}
\begin{eqnarray}
\label{r1}
\sqrt{-\gamma}\gamma^{ab}\tau_{bc}=
\lim_{r \rightarrow \infty} \sqrt{-h} h^{ab}T_{bc}.
\end{eqnarray}
By using this prescription, we find that the stress tensor of the field theory has this simple form
\begin{eqnarray}
\label{st}
\tau^{ab}=\frac{M}{8 \pi l^2}[3 u^a u^b+\gamma^{ab}],
\end{eqnarray}
where $u^a=\delta_t^a$. This is the standard form for the stress tensor of a (2+1) dimensional CFT. 
This tensor is covariantly conserved and manifestly traceless.

For the $k=1$ vacuum Euclidean case, it was conjectured that the strongly coupled dual 
CFT is the infrared limit
of large $N$, $D=3$ super-Yang-Mills theory, 
but now defined on a squashed $S^3$ rather than the round one 
\cite{Emparan:1999pm, Hawking:1998ct}.
However, for the EYM action (\ref{action}) we do not know the underlying boundary CFT.
In particular, we do not know what the $SU(2)$ field corresponds 
to in CFT language.
The bulk YM fields have nothing to do, of course, 
with the nonabelian fields of the dual 
gauge theory.

We end this section with several considerations on the boundary line-element (\ref{l1}), 
not directly related to the aim of this paper.

After the redefinitions $\theta=\rho/l,~l=1/m,~n=\Omega/m^2$, the line element (\ref{l1}) 
can be written  in the form
\begin{equation} 
\label{godel}
d\sigma^2=d\rho^2+\frac{1}{m^2}f_k^2(m \rho)d\varphi^2-\Big(\frac{4\Omega}{m^2}
f_k^2(\frac{m \rho}{2})d\varphi+dt\Big)^2,
\end{equation}
which is the standard form of the nontrivial (2+1)-dimensional part of 
an homogeneous G\"odel-type universe.
This geometry can also be interpreted as resulting from the squashing of three dimensional 
anti-de Sitter (AdS) lightcones \cite{Rooman:1998xf}.

The  general form for a G\"odel-type homogeneous spacetime was found in 1983 
by Rebou\c{c}as and Tiomno \cite{reboucas}. 
It can be expressed as the direct Riemannian sum $dz^2+d\sigma^2$ of a flat factor and the 
$(2+1)$-dimensional metric (\ref{godel}).
The famous G\"odel rotating universe \cite{Godel:1949ga}
corresponds to the case $k=-1,~m^2=2\Omega^2~(l^2=2n^2)$.
The boundary spacetime of a $k=0$ topologically-Taub-NUT-AdS solution corresponds to
the notrivial $(2+1)$-dimensional part of the Som-Raychaudhuri spacetime \cite{Som}.
The value $m^2=4\Omega^2~(l^2=4n^2)$ corresponds to a rotating AdS$_3$ spacetime, 
with the four dimensional
generalization known as the Rebou\c{c}as-Tiomno space-time \cite{Reboucas:wa}. 

The $(2+1)$-dimensional metric (\ref{godel}) contains all the interesting effects exhibited by a 
four-dimensional G\"odel-type spacetime.
We recall that a G\"odel-type model is the archetypal cosmology exhibiting
the properties associated with the rotation of the universe.
Despite the fact that the cosmological rotation is very small 
by the present observation, it cannot be completely ruled out, 
at least in the early universe \cite{Carneiro:2000yw}
(see also \cite{Obukhov:2000jf} for a up to date discussion 
and a large set of references).
Also, the G\"odel model is not the first but perhaps the best known example of a 
solution of Einstein's field equations in which causality 
may be violated and became a paradigm for causality violation 
in gravitational theory.

The properties of the G\"odel-type line elements have been studied by many 
authors. It was found that, for $m^2<4\Omega^2$ they contain CTC  
which cannot be removed by going to a covering space 
and there are no reasonable spacelike surfaces.

Thus asymptotically Taub-NUT-AdS 
solutions provide an opportunity for the study of quantum
field theories on G\"odel-type rotating spacetimes, 
which is clearly an interesting subject \cite{Radu:2001jq}.
Given all the problems with the Lorentzian interpretation of a Taub-NUT-AdS solution, 
we should not be surprised if the boundary metric and the living CFT present unphysical properties.
It is well known \cite{Cassidy:1997kv} that a generic QFT within a causality violating
background exhibit a number of pathologies and we may expect to find 
similar results for a $(2+1)$-dimensional G\"odel-type spacetime.

\section{Concluding remarks}
Even if one's primary interest 
is in asymptotically flat and AdS solutions, 
we hope that, by widening the context to solutions with NUT charge, 
one may achieve a deeper appreciation of the theory. 
There are a number of reasons to consider this type of solutions.
In some supersymmetric theories, closure under duality forces us to consider solutions
with NUT parameter.
Furthermore, dual mass solutions extremize the gravitational action functional in Euclidean 
quantum gravity and therefore cannot be discarded in spite of their causal pathologies.
In particular, one may hopes to attain more general features 
of gravitating nonabelian solutions, 
whether or not containing a magnetic mass.

Our results seem to invalidate a version of the "no hair" conjecture for gravitating 
$SU(2)$ solutions with a magnetic mass, since the structure 
of these solutions is not completely determined by global charges defined at infinity 
such as mass, NUT parameter, magnetic and electric charges.
However, for $\Lambda=0$, even in the presence of a magnetic mass, 
the decay of the electric potential is too rapid to give 
a net electric charge while $\omega(\infty)^2=1$ corresponds to a zero 
magnetic charge. These nonabelian solutions are not dyons in the usual sense but do posses both
electric and magnetic fields.

As expected, the behavior of the solutions with a negative cosmological constant is different
and we find configurations with nonzero nonabelian charges.
We can attribute this behavior to the different 
asymptotic structure of spacetime. 
This point was also crucial in obtaining stable hairy black hole solutions in 
AdS spacetime. 
The presence of a NUT charge does not introduces any new features in this case.
Although such backgrounds are not of great physical interest, they may provide
interesting examples of a more general AdS/CFT correspondence. 

We remark also that, on the Lorentzian section, the mass and NUT parameter are unrelated.
The NUT parameter can be freely specified.
An Euclidean section may be obtained by analytically continuing the coordinate
$t$ and also the parameter $n$.
However, on the Euclidean section, $M$ and $n$ are no longer independent quantities
\cite{Hawking:ig}. 
Not surprisingly, the nonabelian fields in a Taub-NUT instanton background
present a very different behavior as compared to the solutions discussed in this paper.

An important issue that we have not addressed here is the stability of these solutions.
Based on the experience with $n=0, \Lambda<0$ black holes, 
we expect to find that some of the configurations with 
no nodes in $\omega$ are stable solutions against small perturbations.
Also, we have all the reasons to expect that all $\Lambda=0$ solutions are unstable.
The lack of a net electric charge would appear not to improve the chances of stability.

We can as well to include a positive cosmological constant in the theory 
and to look for asymptotically NUT-de Sitter EYM solutions.
Also, a possible extension of this work is to include a dilaton in a theory with $\Lambda=0$.
Based on the experience with $n=0$ EYM-dilaton solutions \cite{Bizon:gi}, we do not expect 
to find any new effects in this case.
However, the situation may be different for a supersymmetric theory. 
An obvious candidate is $N=4,~D=4$ gauged $SU(2)\times SU(2)$ supergravity, also known as 
Freedman-Schwarz model  \cite{Freedman:1978ra}.
Nonabelian BPS exact solutions are known in this case for $n=0$ and a vanishing electric part of the
YM potential \cite{Chamseddine:1997mc}.
It would be interesting to find the corresponding BPS solutions with NUT charge.

Another interesting issue would be to discuss solutions of a 
spontaneously  broken Einstein-Yang-Mills-Higgs theory in the presence of a NUT charge.
It can be proven that in this case
a magnetic monopole necessarily acquires an electric charge.

We did not addressed here the question of inner structure of these solutions (for $r<r_h$).
Given the special global properties of the vacuum Taub-NUT solution \cite{Hawking}, 
this is clearly an interesting subject,
and also the study of the possible effects dual mass could 
have on the singularity structure in a EYM theory.

One may also speculate about the relevance of these results
in string theory and on the possible existence of a duality-like transformation
relating ordinary hairy black holes to  
nonabelian solutions possesing a magnetic mass.

A better understanding of these issues may lead to a deeper understanding of the 
no hair theorems.
\\
\newline
{\bf Acknowledgments} 
\\
The author would like to thank Vladimir Jovanovic for valuable discussions.
This work was performed in the context of the
Graduiertenkolleg of the Deutsche Forschungsgemeinschaft (DFG):
Nichtlineare Differentialgleichungen: Modellierung,Theorie, Numerik, Visualisierung.




\newpage
{\bf Figure Captions}
\\
\\
\\
Figure 1:
\\ 
The gauge potentials $\omega(r)$ and $u(r)$ and the metric functions
$m(r)$ and $e^{-2\delta(r)}$ are plotted as functions of radius 
for typical topological dyon-black hole solutions with $\Lambda<0$. 
\\
\\
Figure 2:
\\ 
The mass-parameter $M$, the values of the electric and magnetic potentials at infinity
$u_0,~\omega_0$ and the electric and magnetic charges $Q_E,~Q_M$ are plotted as a function 
of the value of the magnetic potential on the event horizon $\omega_h$,
for topological black hole solutions with fixed $\Lambda,~r_h,~u_1$. 
Here and in Figures  4 and 5,  $Q_E$ and $Q_M$ are given in units of $V/4\pi$. 
\\
\\
Figure 3:
\\
Typical solutions of the EYM theory with $\Lambda<0$ and a nonvanishing NUT charge. 
\\
\\
Figure 4:
\\
The mass-parameter $M$, the values of the electric and magnetic potentials at infinity
$u_0,~\omega_0$ and the electric and magnetic charges $Q_E,~Q_M$ are plotted as a function of the value of the NUT charge
for $k=1,0,-1$ solutions with fixed $\Lambda,~r_h,~\omega_h,~u_1$.
\\
\\
Figure 5:
\\
The mass-parameter $M$, the values of the electric and magnetic potentials at infinity
$u_0,~\omega_0$ and electric and magnetic charges $Q_E,~Q_M$  
are plotted as a function of the parameter $u_1$ on the horizon,
for $k=1,0,-1$ solutions with fixed $\Lambda,~r_h,~\omega_h,~n$. 
\\
\\
Figure 6:
\\
Typical one- and two-nodes solutions of the EYM theory with NUT charge and $\Lambda=0$. 
The gauge potentials $\omega(r)$ and $u(r)$ and the metric functions
$m(r)$ and $e^{-2\delta(r)}$ are plotted as functions of radius for fixed values of the NUT charge.

\newpage
\setlength{\unitlength}{1cm}

\begin{picture}(16,16)
\centering
\put(-2,0){\epsfig{file=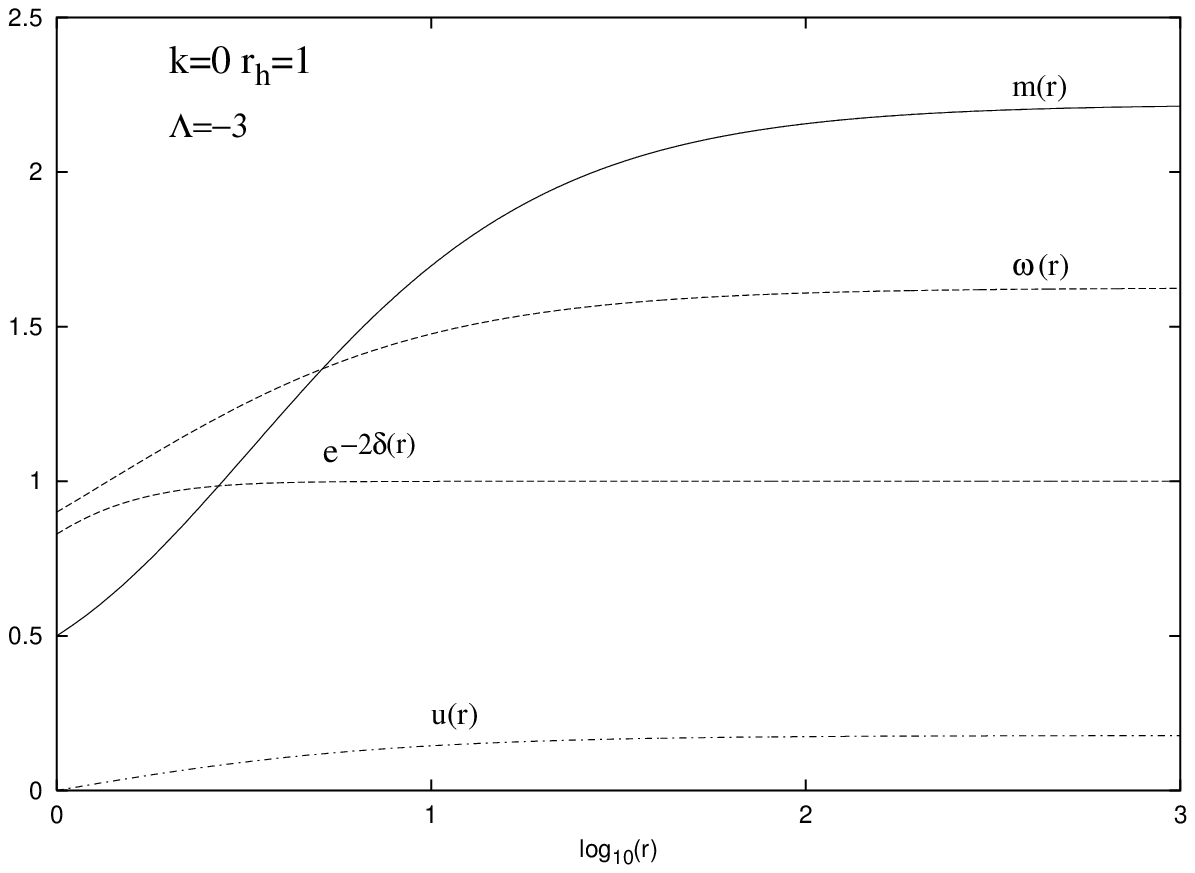,width=16cm}}
\end{picture}
\begin{center}
Figure 1a.
\end{center}

\newpage
\setlength{\unitlength}{1cm}

\begin{picture}(16,16)
\centering
\put(-2,0){\epsfig{file=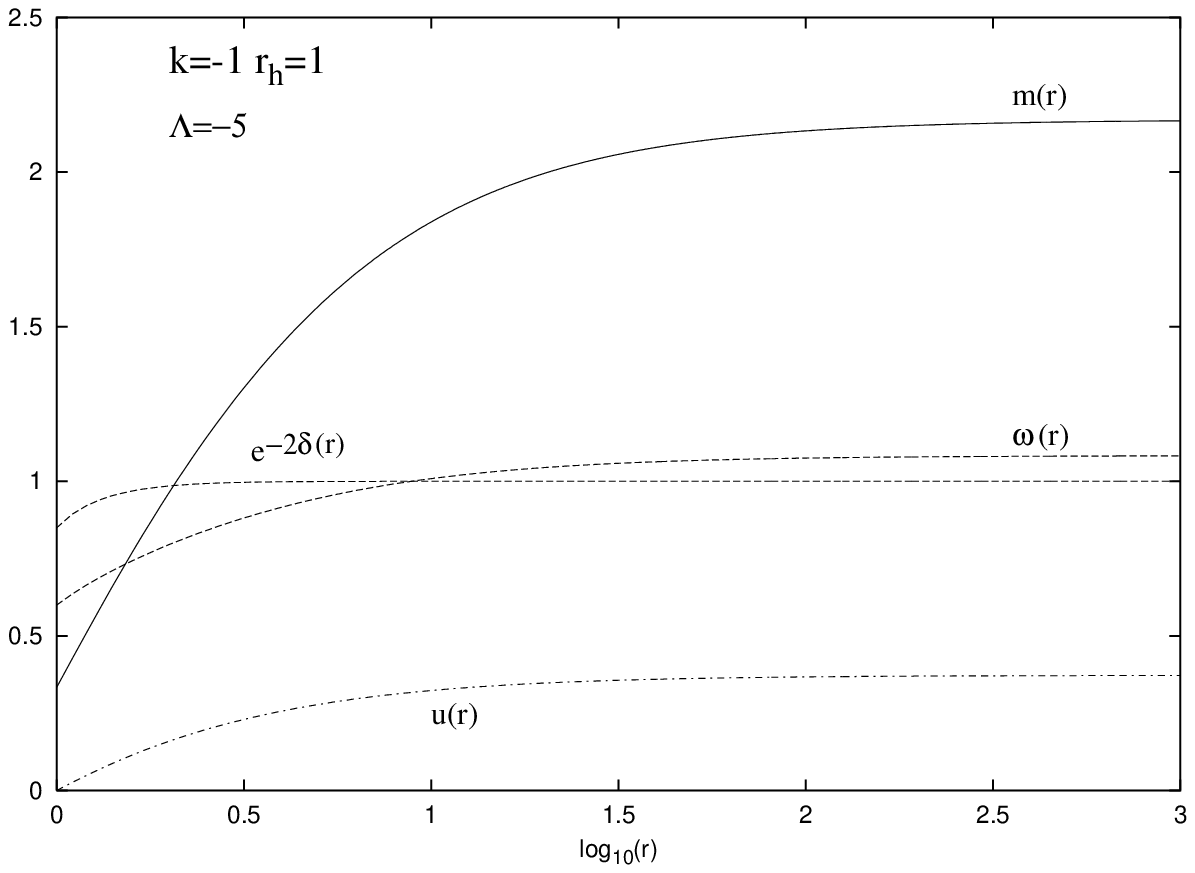,width=16cm}}
\end{picture}
\begin{center}
Figure 1b.
\end{center}

\newpage
\setlength{\unitlength}{1cm}

\begin{picture}(16,16)
\centering
\put(-2,0){\epsfig{file=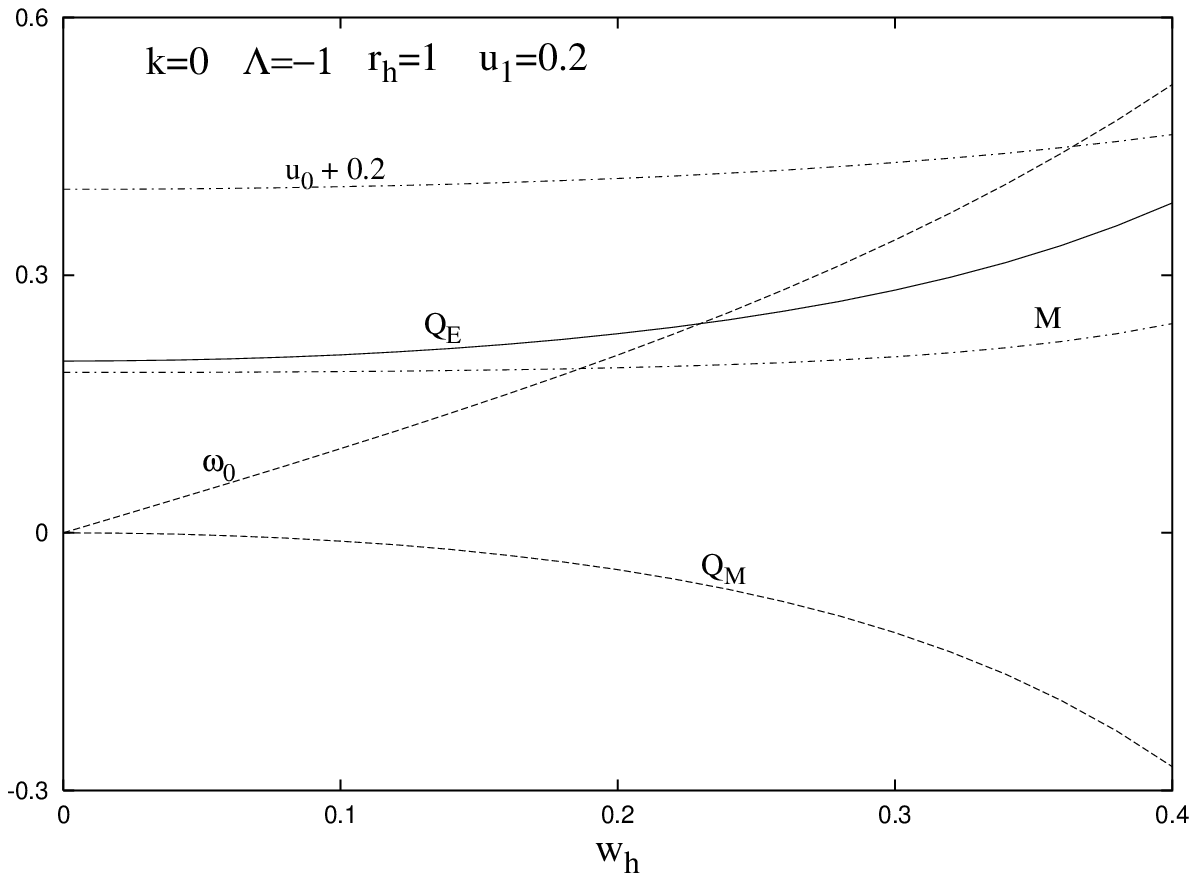,width=16cm}}
\end{picture}
\begin{center}
Figure 2a.
\end{center}

\newpage
\setlength{\unitlength}{1cm}

\begin{picture}(16,16)
\centering
\put(-2,0){\epsfig{file=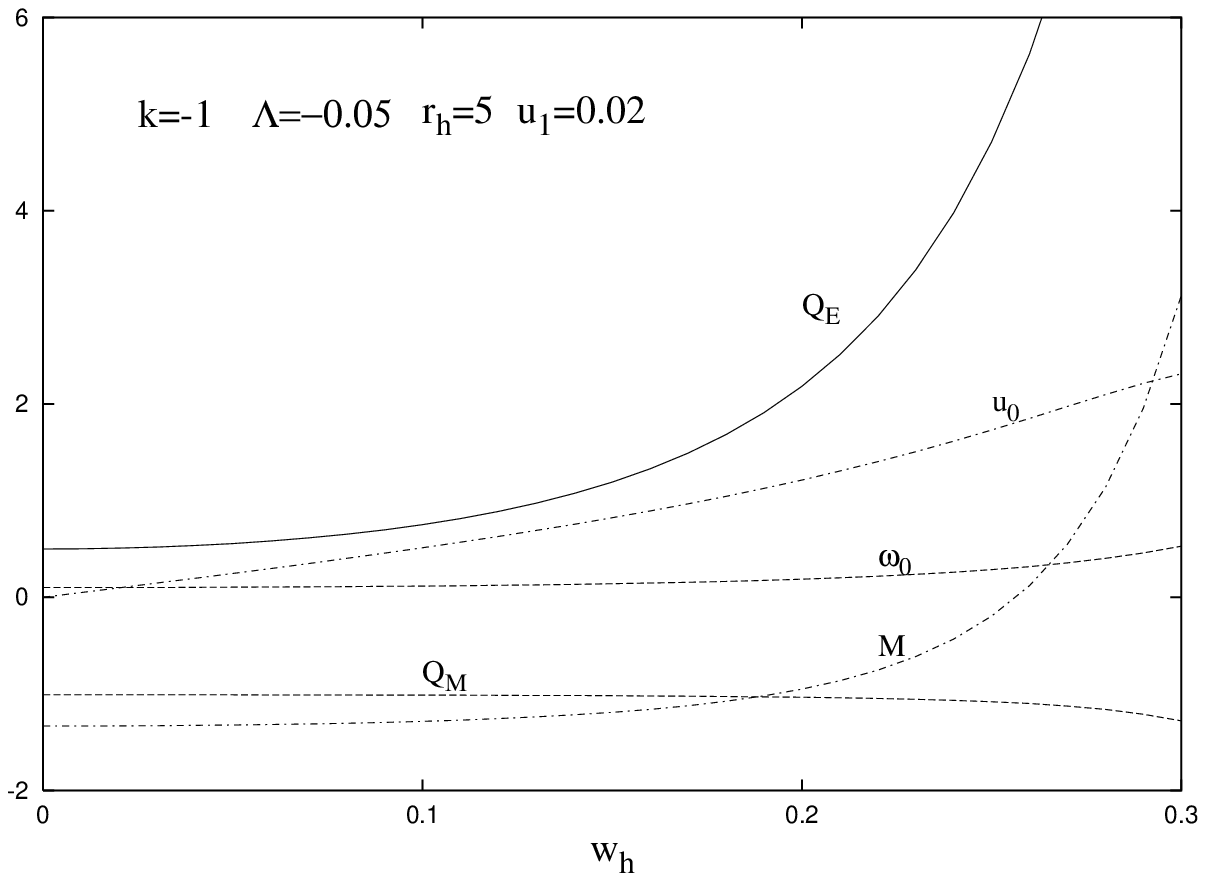,width=16cm}}
\end{picture}
\begin{center}
Figure 2b.
\end{center}

\newpage
\setlength{\unitlength}{1cm}

\begin{picture}(16,16)
\centering
\put(-2,0){\epsfig{file=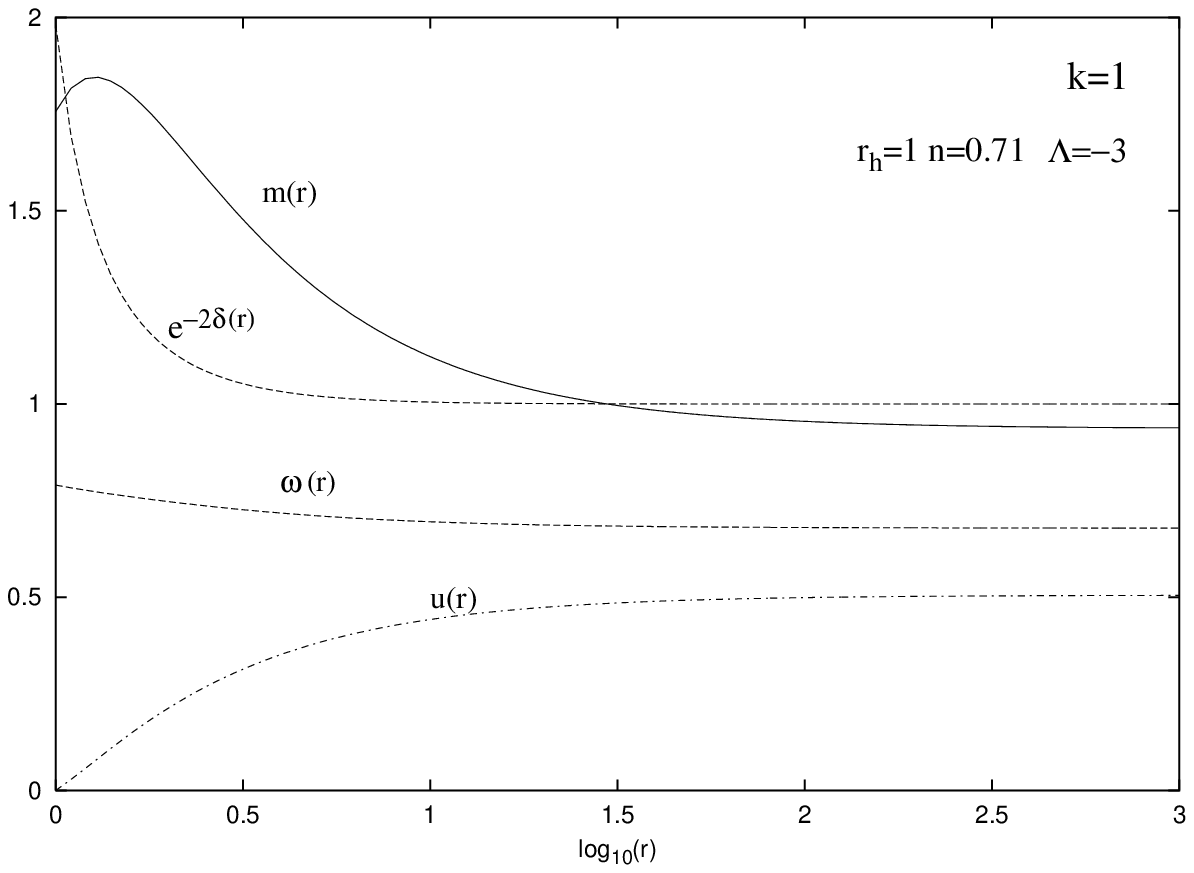,width=16cm}}
\end{picture}
\begin{center}
Figure 3a.
\end{center}

\newpage
\setlength{\unitlength}{1cm}

\begin{picture}(16,16)
\centering
\put(-2,0){\epsfig{file=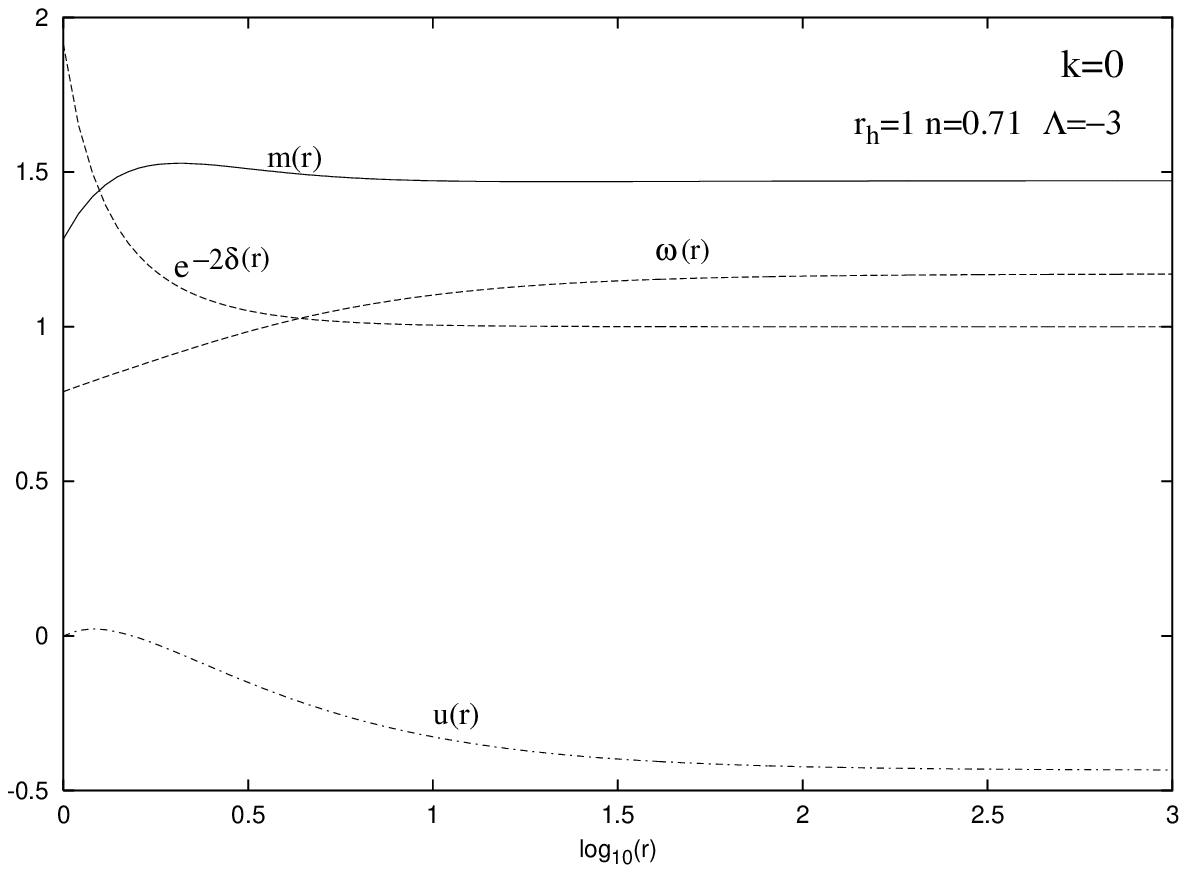,width=16cm}}
\end{picture}
\begin{center}
Figure 3b.
\end{center}

\newpage
\setlength{\unitlength}{1cm}

\begin{picture}(16,16)
\centering
\put(-2,0){\epsfig{file=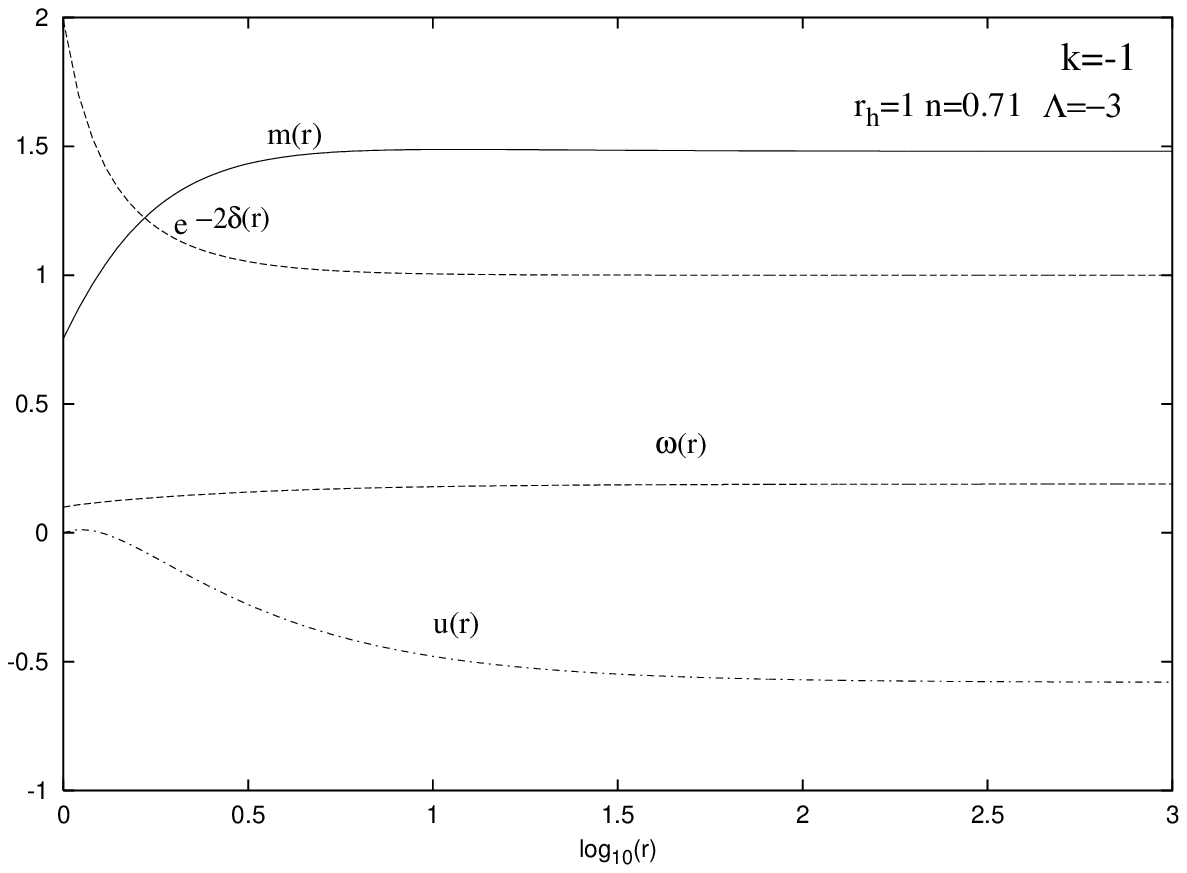,width=16cm}}
\end{picture}
\begin{center}
Figure 3c.
\end{center}

\newpage
\setlength{\unitlength}{1cm}

\begin{picture}(16,16)
\centering
\put(-2,0){\epsfig{file=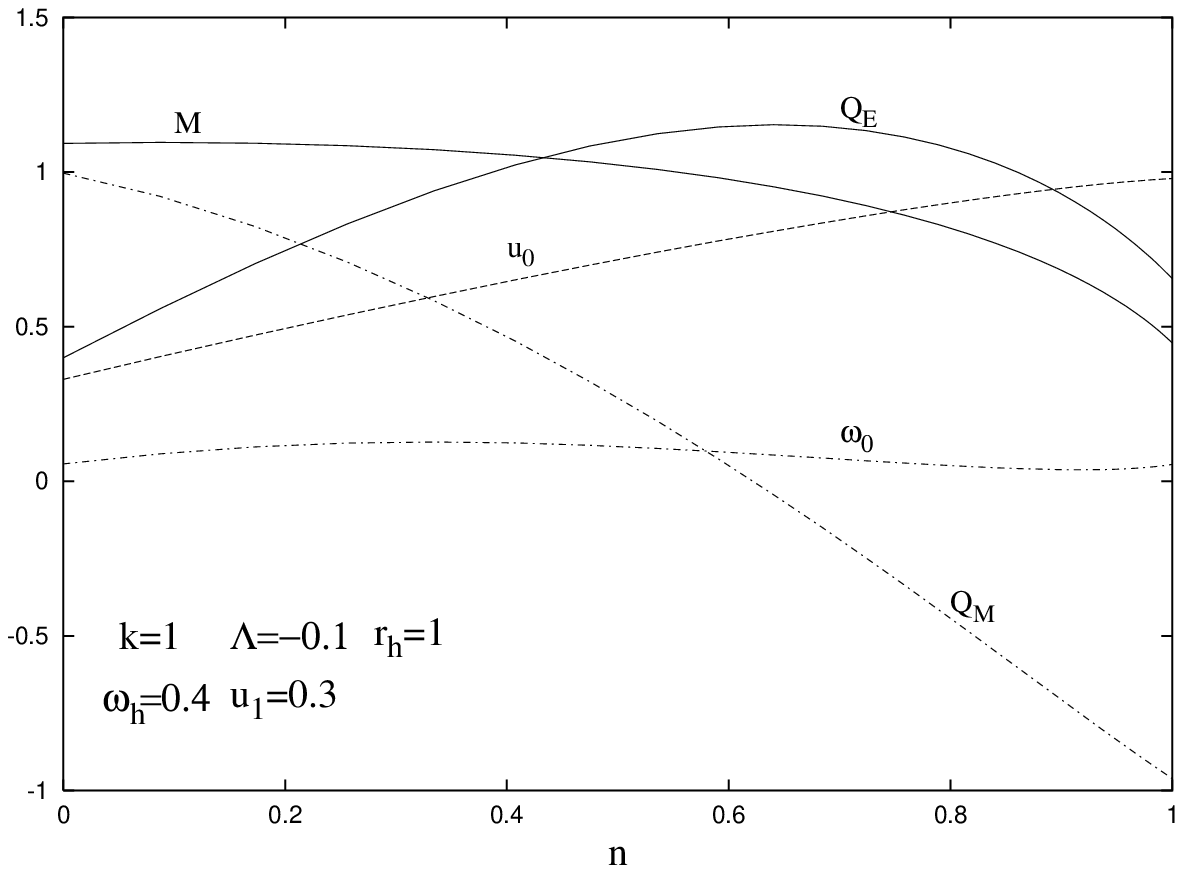,width=16cm}}
\end{picture}
\begin{center}
Figure 4a.
\end{center}
\newpage
\setlength{\unitlength}{1cm}

\begin{picture}(16,16)
\centering
\put(-2,0){\epsfig{file=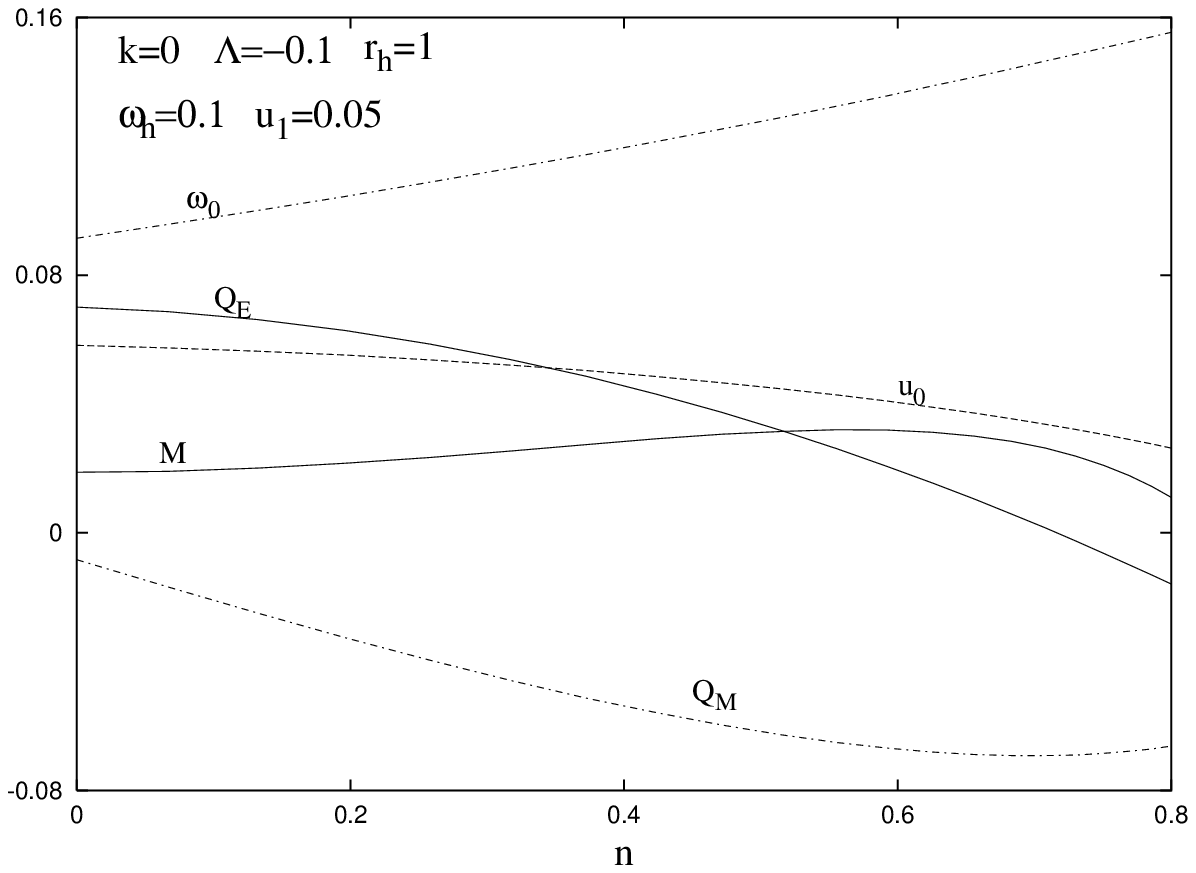,width=16cm}}
\end{picture}
\begin{center}
Figure 4b.
\end{center}

\newpage
\setlength{\unitlength}{1cm}

\begin{picture}(16,16)
\centering
\put(-2,0){\epsfig{file=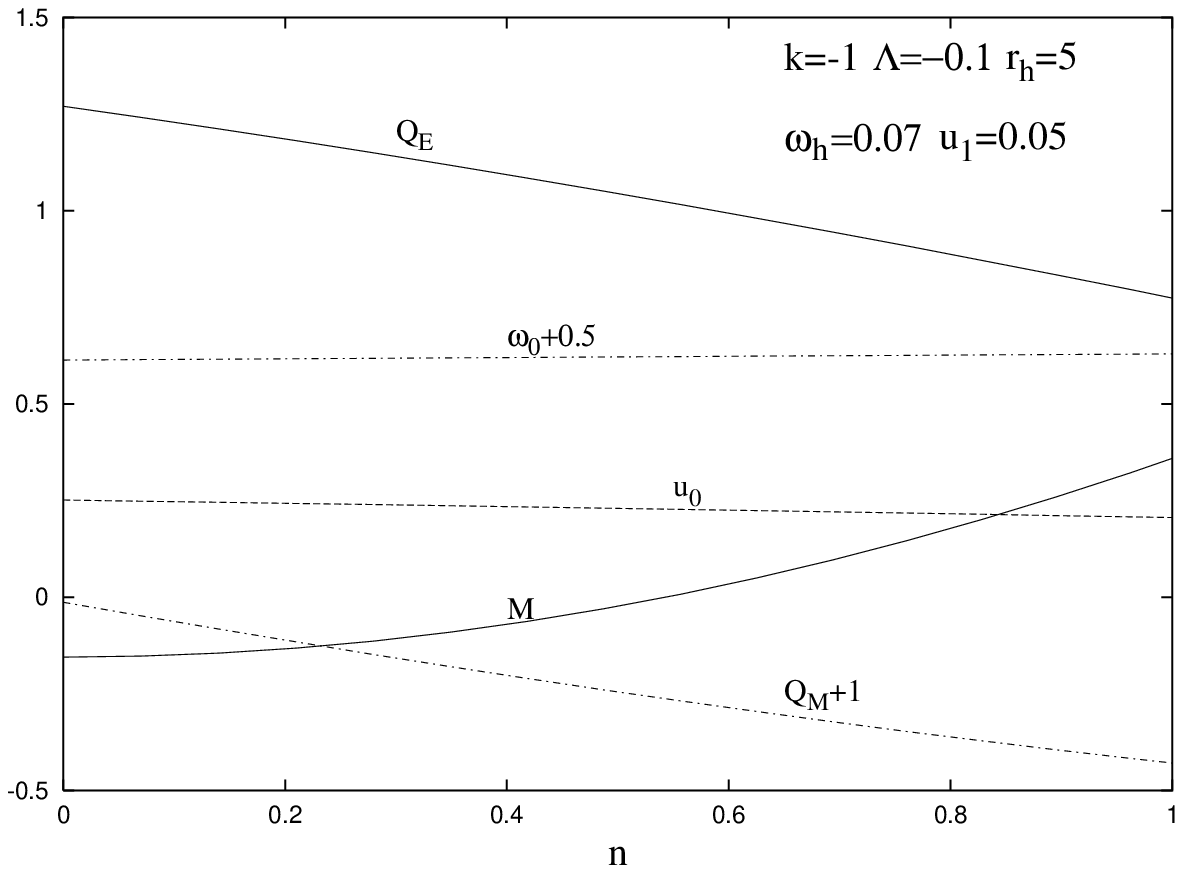,width=16cm}}
\end{picture}
\begin{center}
Figure 4c.
\end{center}

\newpage
\setlength{\unitlength}{1cm}

\begin{picture}(16,16)
\centering
\put(-2,0){\epsfig{file=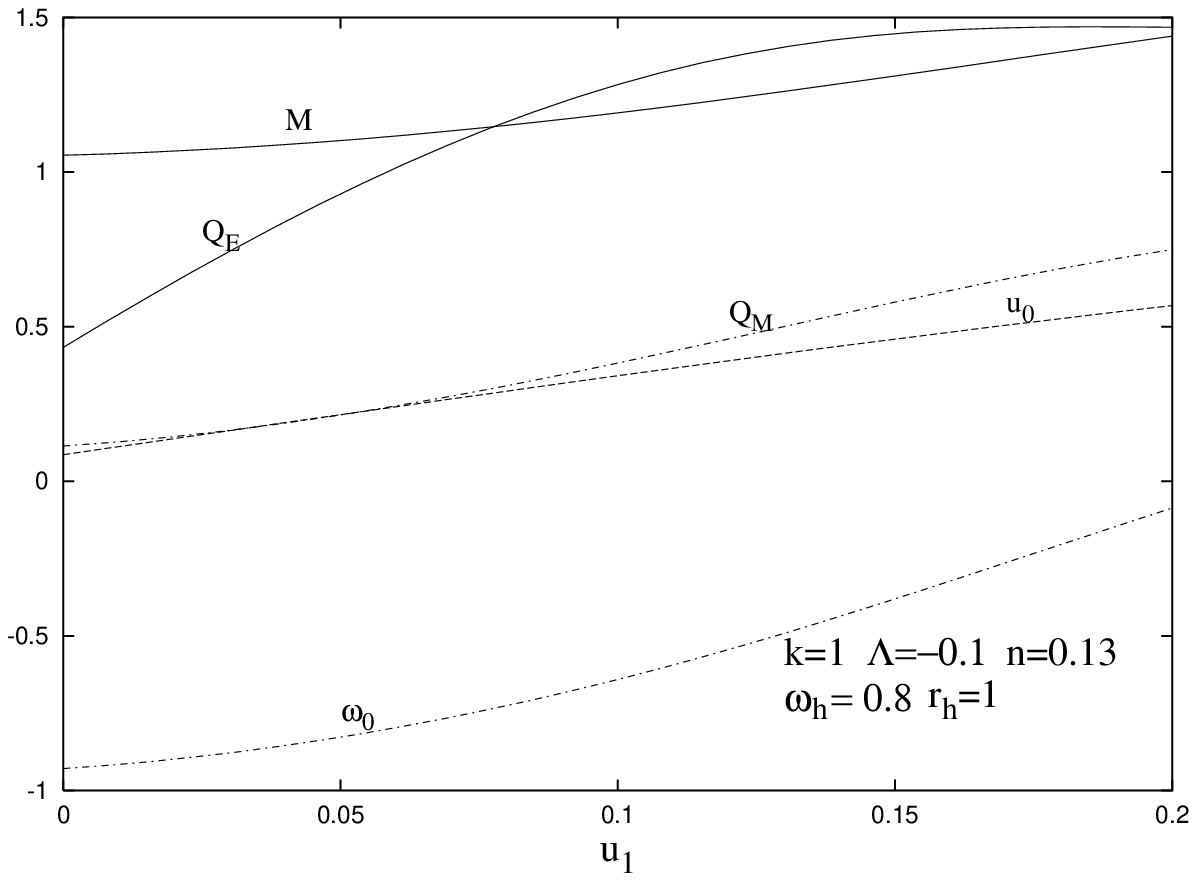,width=16cm}}
\end{picture}
\begin{center}
Figure 5a.
\end{center}

\newpage
\setlength{\unitlength}{1cm}

\begin{picture}(16,16)
\centering
\put(-2,0){\epsfig{file=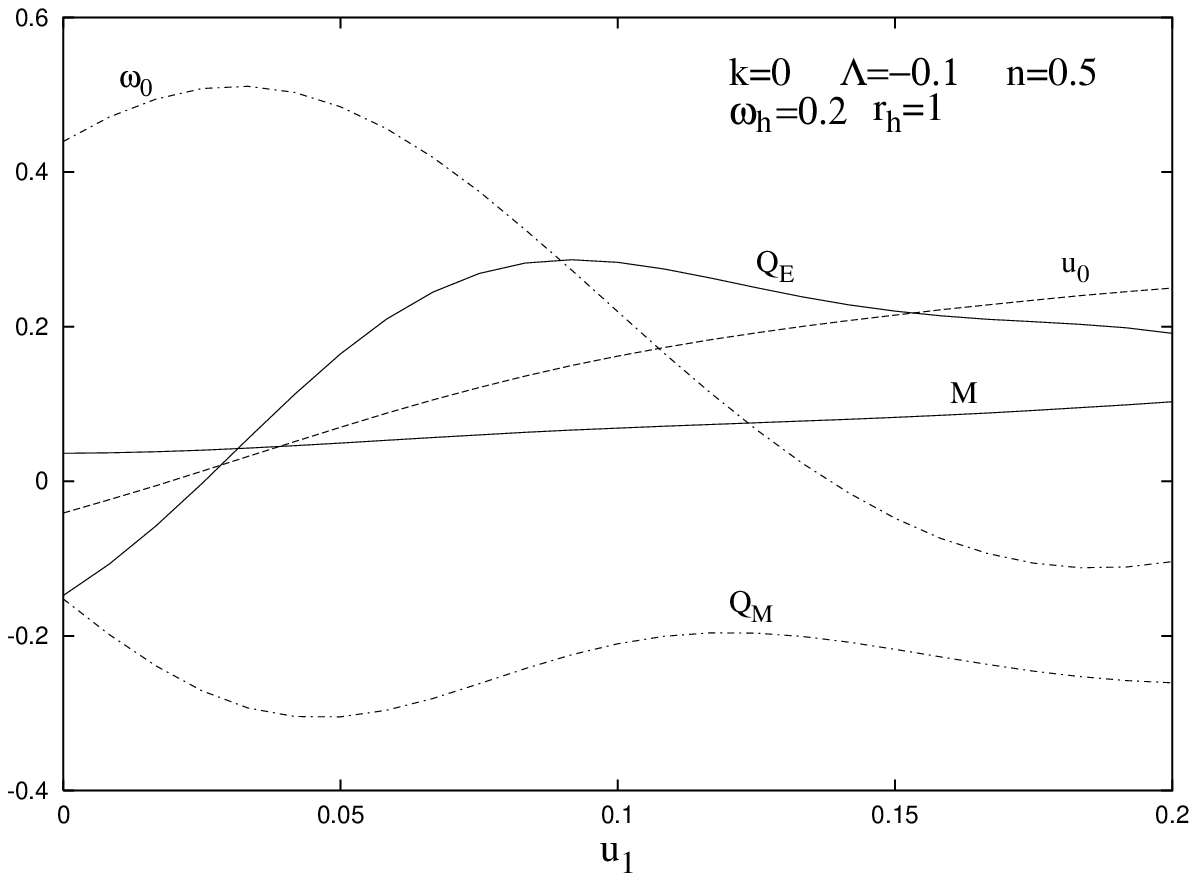,width=16cm}}
\end{picture}
\begin{center}
Figure 5b.
\end{center}

\newpage
\setlength{\unitlength}{1cm}

\begin{picture}(16,16)
\centering
\put(-2,0){\epsfig{file=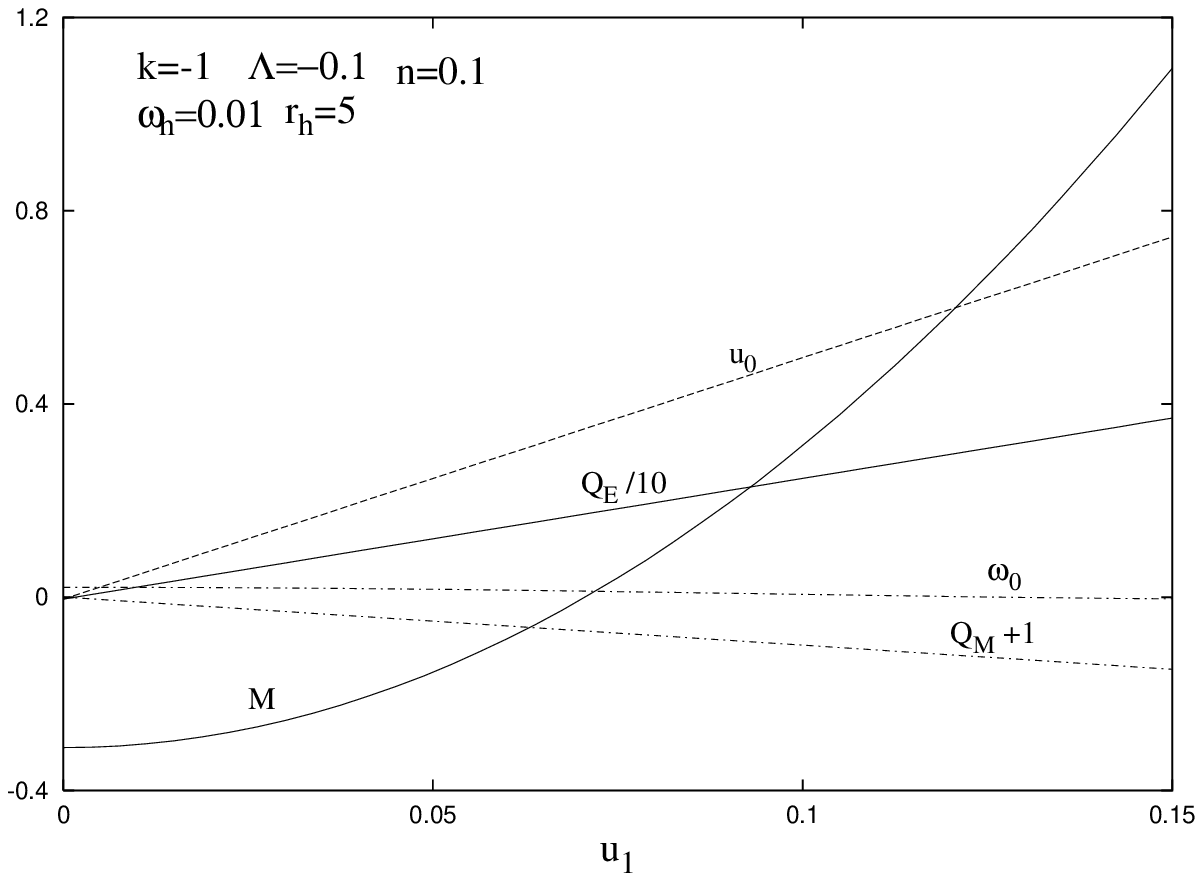,width=16cm}}
\end{picture}
\begin{center}
Figure 5c.
\end{center}
\newpage
\setlength{\unitlength}{1cm}

\begin{picture}(16,16)
\centering
\put(-2,0){\epsfig{file=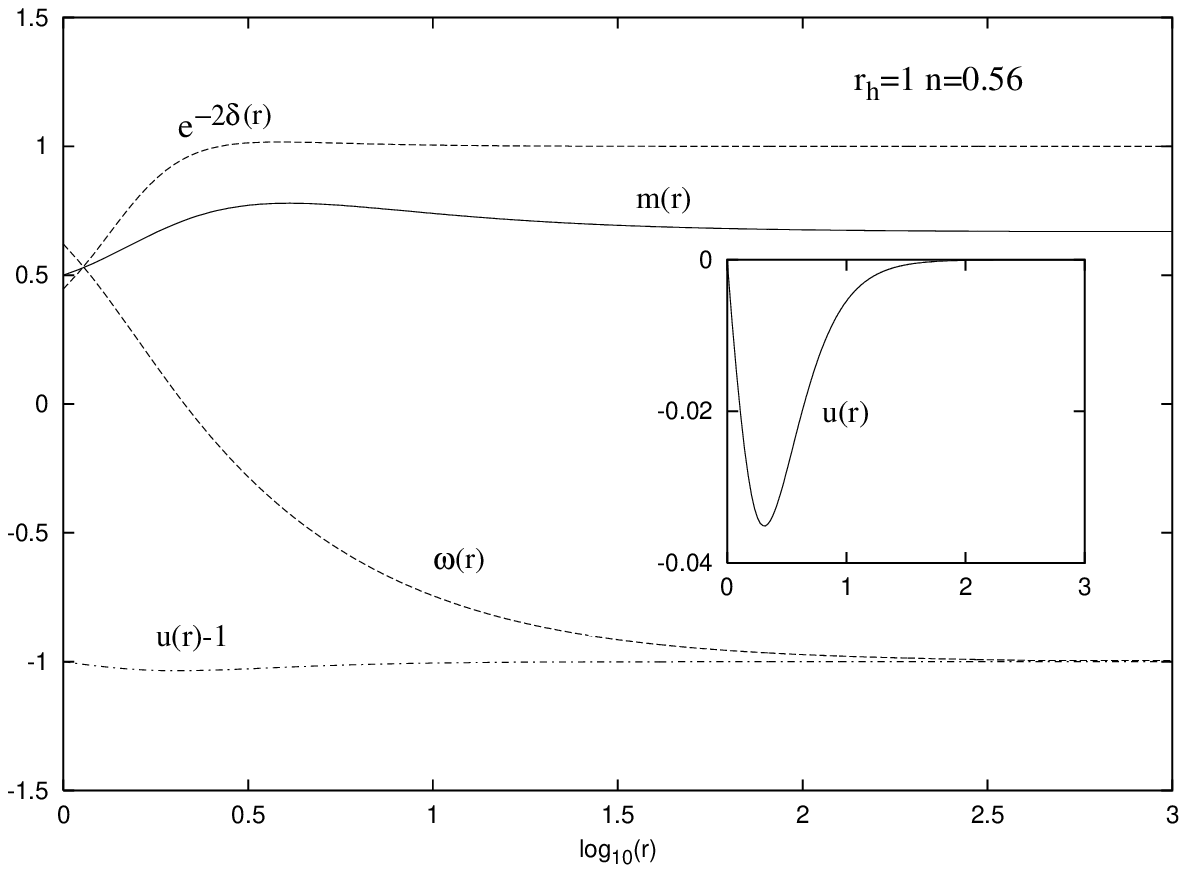,width=16cm}}
\end{picture}
\begin{center}
Figure 6a.
\end{center}

\newpage
\setlength{\unitlength}{1cm}

\begin{picture}(16,16)
\centering
\put(-2,0){\epsfig{file=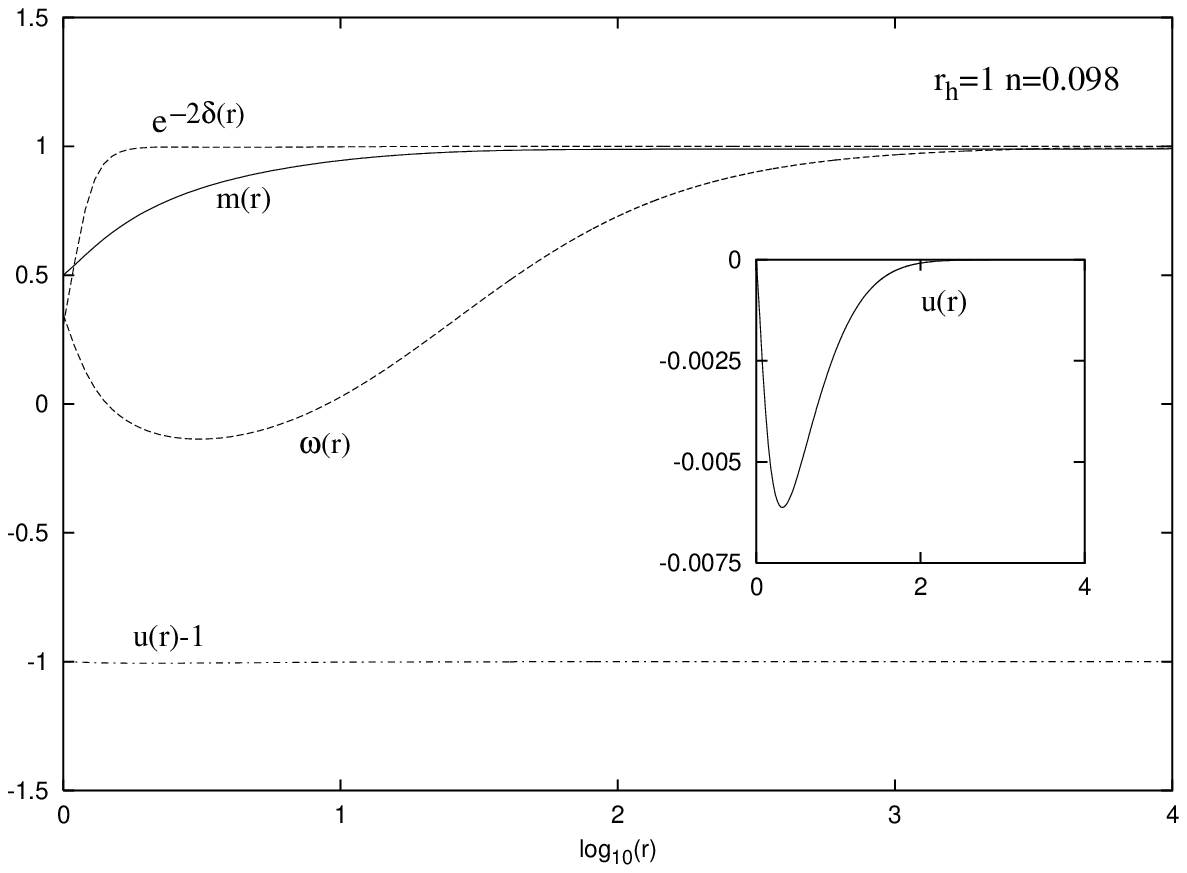,width=16cm}}
\end{picture}
\begin{center}
Figure 6b.
\end{center}

\end{document}